\documentclass[longauth]{aa}

\usepackage{graphicx}
\usepackage{txfonts}
\usepackage{xcolor}
\usepackage{subcaption}
\usepackage{longtable}
\usepackage{multirow}

\usepackage{comment}         
\def\ms{\,m\,s$^{-1}$}         
\def\Mearth{\hbox{$\mathrm{M}_{\oplus}$}}
\def\Rearth{\hbox{$\mathrm{R}_{\oplus}$}}
\def\deg{\hbox{$^\circ$}}

\def\Rsun{\,R$_{\odot}$}
\def\Msun{\,M$_{\odot}$}    
\def\gcm{\,g\,cm$^{-3}$}

\def\ms{\,m\,s$^{-1}$}         
\def\Mearth{\hbox{$\mathrm{M}_{\oplus}$}}
\def\Rearth{\hbox{$\mathrm{R}_{\oplus}$}}
\def\deg{\hbox{$^\circ$}}

\def\Rsun{\,R$_{\odot}$}
\def\Msun{\,M$_{\odot}$}    
\def\gcm{\,g\,cm$^{-3}$}

\usepackage{natbib,twoopt}
\usepackage[breaklinks=true]{hyperref} 
\bibpunct{(}{)}{;}{a}{}{,}             
\makeatletter
  \newcommandtwoopt{\citeads}[3][][]{\href{http://adsabs.harvard.edu/abs/#3}%
    {\def\hyper@linkstart##1##2{}%
     \let\hyper@linkend\@empty\citealp[#1][#2]{#3}}}
  \newcommandtwoopt{\citepads}[3][][]{\href{http://adsabs.harvard.edu/abs/#3}%
    {\def\hyper@linkstart##1##2{}%
     \let\hyper@linkend\@empty\citep[#1][#2]{#3}}}
  \newcommandtwoopt{\citetads}[3][][]{\href{http://adsabs.harvard.edu/abs/#3}%
    {\def\hyper@linkstart##1##2{}%
     \let\hyper@linkend\@empty\citet[#1][#2]{#3}}}
  \newcommandtwoopt{\citeyearads}[3][][]%
    {\href{http://adsabs.harvard.edu/abs/#3}
    {\def\hyper@linkstart##1##2{}%
     \let\hyper@linkend\@empty\citeyear[#1][#2]{#3}}}
\makeatother

\usepackage{color}
\hypersetup{colorlinks=true,linkcolor=blue,citecolor=blue,urlcolor=blue}

\usepackage[breaklinks=true]{hyperref} 

\begin{document} 

   \title{A decade of monitoring the HIP~41378 planetary system} \subtitle{Masses and orbital periods of six planets and a planet candidate}

\author{S.~Grouffal\inst{\ref{LAM},\ref{IPAG}}
          \and
          A.~Santerne\inst{\ref{LAM},\ref{IPAG}}
          \and
          X.~Dumusque\inst{\ref{Geneva}}
          \and
          B.~Akinsanmi\inst{\ref{Geneva}}
          \and
          T.~Guillot\inst{\ref{Nice}}
          \and
          N.~C.~Hara\inst{\ref{LAM}}
          \and
          A.~Leleu\inst{\ref{Geneva}}
          \and
          L.~Malavolta\inst{\ref{Padova1},\ref{Padova2}}
          \and
          M.~Saillenfest\inst{\ref{LTE}}
          \and
          D.~J.~Armstrong\inst{\ref{Warwick1},\ref{Warwick2}}
          \and
          S.~C.~C.~Barros\inst{\ref{caup},\ref{uPorto}}
          \and 
          D.~Bayliss\inst{\ref{Warwick1}}
          \and
          A.~S.~Bonomo\inst{\ref{Torino}}
          \and
          D.~J.~A.~Brown\inst{\ref{Warwick1}, \ref{Warwick2}}
          \and
          A. Collier Cameron\inst{\ref{StAndrews}}
          \and
          M. Cretignier\inst{\ref{Oxford}}
          \and
          I.~J.~M.~Crossfield\inst{\ref{UKansas}}
          \and
          F.~Dai\inst{\ref{hawaii}}
          \and
          M.~Damasso\inst{\ref{Torino}}
          \and
          O.~Demangeon\inst{\ref{caup}}
          \and
          P.~Figueira\inst{\ref{Granada}}
          \and
          P.~Leonardi\inst{\ref{Padova1}}
          \and
          A. F. Martínez Fiorenzano\inst{\ref{Tenerife}}
          \and
          M.~L\'opez-Morales\inst{\ref{stsci}}
          \and
          E.~Molinari\inst{\ref{Brena}}
          \and
          A.~Mortier\inst{\ref{Birmingham}}
          \and
          L.~D.~Nielsen\inst{\ref{Munchen}}
          \and
          H.~P.~Osborn\inst{\ref{Bern},\ref{Zurich}}
          \and
          E.~Petigura\inst{\ref{UCLA}}
          \and
          K.~Rice\inst{\ref{Edinburgh1}, \ref{Edinburgh2}}
          \and
          N.~C.~Santos\inst{\ref{caup},\ref{uPorto}}
          \and 
          A.~Sozzetti\inst{\ref{Torino}}
          \and
          S.~Sulis\inst{\ref{LAM}}
          \and
          S.~Udry\inst{\ref{Geneva}}
          \and
          C.~Watson\inst{\ref{Belfast}}
          }

   \institute{Aix Marseille Univ, CNRS, CNES, Institut Origines, LAM, Marseille, France\label{LAM}
         \and
             Univ. Grenoble Alpes, CNRS, IPAG, 38000 Grenoble, France\label{IPAG}
        \and
            Department of Astronomy, University of Geneva\label{Geneva}
        \and 
            Université Côte d’Azur, Observatoire de la Côte d’Azur, CNRS, Laboratoire Lagrange, CS 34229, F-06304 Nice Cedex 4,France \label{Nice} 
        \and
            Dipartimento di Fisica e Astronomia, Universit\`a degli Studi di Padova, Vicolo dell’Osservatorio 3, 35122 Padova, Italy\label{Padova1}
        \and 
            LTE, Observatoire de Paris, Universit{\'e} PSL, Sorbonne Universit{\'e}, Universit{\'e} de Lille, LNE, CNRS, 75014 Paris, France\label{LTE} 
        \and
            INAF, Osservatorio Astronomico di Padova, Vicolo dell’Osservatorio 5, 35122 Padova, Italy\label{Padova2}
        \and
            Department of Physics, University of Warwick, Gibbet Hill Road, Coventry, CV4 7AL, UK\label{Warwick1}
        \and
            Sub-department of Astrophysics, Department of Physics, University of Oxford, Oxford OX1 3RH, UK\label{Oxford}
        \and
            Department of Physics and Astronomy, University of Kansas, Lawrence, KS, USA\label{UKansas}
        \and
            Centre for Exoplanets and Habitability, University of Warwick, Gibbet Hill Road, Coventry, CV4 7AL, UK\label{Warwick2}
        \and
            INAF - Osservatorio Astronomico di Brera\label{Brena}
        \and
            University Observatory, Faculty of Physics, Ludwig-Maximilians-Universit{\"a}t M{\"u}nchen, Scheinerstr. 1, 81679 Munich, Germany\label{Munchen}
        \and
            Center for Space \& Habitability, Physikalisches Institut, Universit\"at Bern, Gesellschaftsstrasse 6, 3012 Bern, Switzerland\label{Bern}
        \and
            Inst. f. Teilchen- und Astrophysik, ETH Z\"urich, Wolfgang-Pauli-Strasse 27, 8093 Z\"urich, Switzerland\label{Zurich}
        \and
            Space Telescope Science Institute, 3700 San Martin Drive, Baltimore MD 20218, USA\label{stsci}
        \and
            Instituto de Astrof\'isica e Ci\^encias do Espa\c{c}o, Universidade do Porto, CAUP, Rua das Estrelas, 4150-762 Porto, Portugal\label{caup}
        \and
            Departamento de F\'isica e Astronomia, Faculdade de Ci\^encias, Universidade do Porto, Rua do Campo Alegre, 4169-007 Porto, Portugal\label{uPorto}
        \and
            Institute for Astronomy, University of Hawai`i, 2680 Woodlawn Drive, Honolulu, HI 96822, USA\label{hawaii}
        \and
            SUPA, Institute for Astronomy, University of Edinburgh, The Royal Observatory, Blackford Hill, Edinburgh EH9 3HJ, UK\label{Edinburgh1}
        \and 
            Centre for Exoplanet Science, University of Edinburgh, Edinburgh EH9 3HJ, UK\label{Edinburgh2}
        \and
            INAF - Osservatorio Astrofisico di Torino, via Osservatorio 20, 10025 Pino Torinese, Italy\label{Torino}
        \and
            SUPA School of Physics \& Astronomy, University of St Andrews, North Haugh, St Andrews, UK\label{StAndrews}
        \and
            Department of Physics \& Astronomy, University of California Los Angeles, Los Angeles, CA 90095, USA\label{UCLA}
        \and
            School of Physics \& Astronomy, University of Birmingham, Edgbaston, Birmingham, B15 2TT, UK\label{Birmingham}
        \and
            Instituto de Astrof\'{i}sica de Andaluc\'{i}a-CSIC, Glorieta de la Astronom\'{i}a s/n, E-18008 Granada, Spain\label{Granada}
        \and
         Fundación Galileo Galilei - INAF, Rambla José Ana Fernandez Pérez 7, E-38712 Breña Baja, Tenerife, Spain \label{Tenerife}
        \and 
            Astrophysics Research Centre, Queen’s University Belfast, Belfast, BT7 1NN, UK\label{Belfast}}

   \date{Received 3 March 2026; accepted 1 April 2026}

  \abstract
   {Multi-planetary systems provide key constraints on planet formation and evolution, as their architecture encodes the dynamical history of planets formed within a common protoplanetary disc. However, the current population remains strongly biased towards compact, short-period systems, and only a limited number of such systems with measured masses and radii are known. 
   HIP~41378 is an exceptional system hosting five transiting planets with orbital periods of up to 1.5 years, including an ultra-low-density planet, HIP~41378~f. The outer transiting planets d and e remained poorly constrained with unknown periods and masses, leaving the system architecture only partially characterised.  We present long-term monitoring of this target with high-precision radial-velocity (RV) instruments (HARPS, HARPS-N, HIRES, and ESPRESSO) and space-based photometry spanning 2015-2024. We detected RV signals for all the planets, confirming their orbital periods and constraining their masses. In particular, the RV data strongly favour an orbital period of $P_d = 278$ days for planet d  and refine the orbital period of planet e to $P_e = 393_{-5}^{+3}$ days. We measured a new mass of $M_f = 25 \pm 5$ \Mearth\ for HIP~41378~f, confirming its super-puff nature with a bulk density of $\rho_f = 0.166^{+0.033}_{-0.036}$\gcm. We also confirm the planetary nature of HIP~41378~g, a non-transiting planet with a 63-day period, and determine its minimum mass. In addition, the RVs reveal a long-period signal, with $P = 2602_{-433}^{+468}$ days, which we attribute to the candidate planet HIP~41378 h, although a stellar magnetic cycle cannot be excluded.
    Finally, we investigated the system’s dynamical architecture and resonant structure, assessed its completeness by constraining additional undetected planets, and looked into the implications for the origin and internal structure of the remarkable planet HIP 41378 f.}
    
   \maketitle

\section{Introduction}

Multi-planetary systems are of high interest in planetology. Beyond the question of whether the Solar System is unique or not, multiple extra-solar systems exhibit a large diversity \citep[e.g.][]{Lissauer2014,Muresan2024}, providing us with new insights into the processes that govern planetary formation and evolution \citep[e.g.][]{Ford2014,Mishra2023}. Moreover, they enable direct comparative studies both within a single system \citep[e.g.][]{Ciardi2013, Millholland2017, Weiss2018, Otegi2022} and between systems \citep[e.g.][]{Alibert2019, Gilbert2020, Mishra2023}. Within a given system, the planets formed contemporaneously from a common proto-planetary disc, although at different locations. Their present-day properties may have preserved imprints of their common dynamical history and formation pathways, albeit with a different evolution.

As of February 2026, more than 1026 planetary systems with at least two confirmed planets have been detected.\footnote{From  NASA Exoplanet Archive; consulted in February 2026.} These systems provide valuable constraints on a wide range of planetary parameters such as masses, radii, orbital spacing, eccentricities \citep[e.g.][]{VanEylen2015}, and dynamical stability \citep[e.g.][]{Laskar2017, Stalport2022}.
However, the current observational picture of transiting planetary system architectures remains strongly affected by observational biases. Transit surveys such as \textit{Kepler} \citep{Borucki2010} are more sensitive to compact, short-period systems, while planets on wider orbits remain poorly explored due to their low geometric probability. Combining observational results with population synthesis models has led to important theoretical predictions that now require confirmation by upcoming observations \citep[e.g.][]{Mordasini2009, Emsenhuber2021}. The Transiting Exoplanet Survey Satellite (TESS) survey expands on the \textit{Kepler} sample, but still primarily detects short-period planets (P $\leq$ 50 days). As of today, no theoretical model is able to accurately reproduce the observed population of multi-planetary systems \citep[e.g.][]{Turtelboom2025}.

One intriguing pattern identified in many \textit{Kepler} systems is the so-called peas-in-a-pod configuration, in which planets are found to be similar in size and evenly spaced in terms of orbital distance \citep{Weiss2018}. This architecture has been interpreted as the outcome of smooth and synchronous formation, followed by convergent migration and resonance trapping within the protoplanetary disc \citep{Izidoro2017, Goldberg2022}. Systems such as TRAPPIST-1 \citep{Gillon2017} and Kepler-223 \citep{Mills2016} are typical of this architecture. However, recent studies have questioned the universality of this trend \citep[e.g.][]{Murchikova2020, Zhu2020}, suggesting that additional observations are required to determine how common it truly is. Notably, \citet{Millholland2022} reported a possible breakdown of the pattern for planets with orbital periods of between 100 and 300 days. Expanding the sample of systems with longer period exoplanets is therefore crucial to test theories of formation and evolution. 

In this context, HIP~41378 is a unique planetary system; this bright star hosts five transiting planets with long orbital periods, as revealed by the \textit{K2} mission \citep{Vanderburg2016}. The system is composed of three inner planets named b, c, and g (P$_b$ = 15 days, P$_c$ = 31 days, P$_g$ $\approx$ 64 days) \citep{Leonardi2025}; and three outer planets named d, e, and f, with orbital periods of up to 1.5 years. While planets b and c are mini-Neptunes, planets d, e, and f are larger, reaching up to the size of Saturn. Planet g is not transiting. In this system, HIP~41378~f is of particular interest: with a period of 542 days (T$_{eq}$ $\approx$ 294 K ) and a radius of $9.2 \pm 0.1$ \Rearth, it stands out as an extremely low-density 'super-puff' planet with a 95$\%$ credible upper limit on its density of 0.13 \gcm \citep{Santerne2019}. The orbital periods and masses of HIP~41378 d and e remain poorly constrained \citep{Grouffal2022, Sulis2024}. Determining the architecture and compositions of these three outer planets is essential for a complete understanding of the system.

In this paper, we present a global characterisation of the HIP~41378 planetary system based on radial-velocity (RV) observations obtained with the HARPS, HARPS-N, HIRES, and ESPRESSO spectrographs between 2016 and 2025. The goals of this work are to measure the masses of all planets, refine the orbital periods of the long-period planets, and place the system into context with other multi-planetary architectures. HIP 41378 thus serves as an ideal testbed for future investigations of temperate, multi-planet systems. An overview of the system and a summary of previous results are presented in Section \ref{sec:overview}. Then, the observations are described in Section \ref{sec:obs}. In Section \ref{sec:analyse_RV}, we analyse the RVs and describe our investigation into the planet's orbital periods and stellar activity. In Section \ref{sec:transit_RV}, we present a joint analysis of the photometry and RVs. Finally, we discuss the possible orbital periods of the outer planets, the nature of HIP~41378 f, and the dynamics of the system in Section \ref{sec:discussion}. The conclusions are given in Section \ref{sec:conclusion}.

\section{System overview}
\label{sec:overview}

The system around HIP~41378 (V-mag = 8.93) has been the subject of numerous studies, which have progressively refined our understanding of its architecture. Five transiting planets were discovered in the 70-day campaign C5 of the \textit{K2} space mission \citep{Vanderburg2016}, and four of them were re-observed in transit during the C18 campaign in 2018 \citep{Berardo2019, Becker2019}. The two inner sub-Neptunes (planets b and c, with orbital periods of 15.6 and 31.7 days) have been observed in transits during the two K2 campaigns, as well as in eight sectors of TESS and subsequent observations with CHEOPS \citep{Leonardi2025}. 
The outer part of the system hosts three transiting planets, named HIP~41378 d, e, and f. Planets d and f were observed once in each K2 campaign, leading to a range of possible orbital periods. Since planet e was observed only in the first campaign, its orbital period is unknown. 
The characterisation of the system improved through the asteroseismology study of the star. \citet{Lund2019} used the K2 C18 short cadence data for the asterosismic study of the star and derived precise stellar parameters: a stellar radius of $1.300 \pm 0.009 $ \Rsun, a stellar mass of $1.22^{+0.03}_{-0.02}$ \Msun, and an age of $2.07^{+0.36}_{-0.27}$ Gyr (see Table \ref{tab:StellarParameters}). These precise stellar parameters, combined with transit durations of planets d, e, and f, allowed for improved estimates of their orbital periods, all exceeding 200 days. The period of planet f was further confirmed at 542 days by the first RV analysis \citet{Santerne2019}.

\begin{table}[]
    \caption{List of stellar parameters.}
    \centering
    \begin{tabular}{lcc}
    \hline \hline
    Parameter & Value & Source \\
    \hline
    Name & HIP41378 & (1)\\
    Right Ascension$^a$ & 08h26m27.85s & (1)\\
    Declination$^a$ & +10d04m49.334s & (1)\\
    V magnitude & 8.901 $\pm$ 0.096 & (2)\\
    H magnitude & 7.768 $\pm$ 0.038 & (3)\\
    RA proper motion [mas.$y^{-1}$] & -48.0020 $\pm$ 0.0203 & (6)\\
    Parallax [mas] & 9.4360 $\pm$ 0.0208 & (6)\\
    \hline
    Effective temperature [K] & 6290 $\pm$ 77 & (4)\\
    Iron abundance [dex] & $-$0.05 $\pm$ 0.10 & (4)\\
    Surface gravity [cgs] & 4.298 $\pm$ 0.004 & (4)\\
    Stellar density [\gcm] & 0.785 $\pm$ 0.008 & (4)\\
    Stellar mass [\Msun] & 1.22$^{_{+0.03}}_{^{-0.02}}$ & (4)\\
    Stellar radius [\Rsun] & 1.300 $\pm$ 0.009 & (4)\\
    Stellar age [Gyr] & 2.07$^{_{+0.36}}_{^{-0.27}}$ & (4)\\
    Distance [pc] & 105.60$_{-0.17}^{+0.24}$ & (5)\\
    \hline
    \end{tabular}
    \tablefoot{$^a$ ICRS -- J2000}
    \tablebib{(1) \cite{Wenger2000}; (2) \citet{Droege2006}; (3) \citet{Cutri2003}; (4) \citet{Lund2019}; (5) \citet{BailerJones2021}; (6) \citet{Gaia2023}.}
    \label{tab:StellarParameters}
\end{table}

 The planet HIP~41378~f has drawn considerable attention due to its extremely low density, which remains unexplained by current theories \citep[e.g.][]{Alam2022, Belkovski2022}. Six transits of this planet have been observed to date: two with K2, one from the ground with NGTS \citep{Bryant2021}, one in transmission spectroscopy with the Hubble Space Telescope (HST) \citep{Alam2022}, one via the Rossiter–McLaughlin effect \citep{Grouffal2025}, and an additional event from the ground with Tierras, TRAPPIST-North, and LCOGT \citep{Garcia-Mejia2025}. Its orbital period has been measured at 542.07 days, and its radius is $R_f = 9.2 \pm 0.1$ \Rearth. RV monitoring by \citet{Santerne2019} constrained its mass to $M_p = 12 \pm 3$~\Mearth, corresponding to an extraordinarily low bulk density of $\rho_f = 0.09 \pm 0.02$\gcm\ for such a temperate, long-period, and mature planet. Several hypotheses have been proposed to explain this super-puff \citep{Lee2016} nature, including the presence of circumplanetary rings \citep{Akinsanmi2020} or high-altitude photochemical hazes \citep{Wang2019,Gao2020}. \citet{Belkovski2022} attempted to model HIP~41378~f’s structure with H/He envelope models \citep{Lopez2012} and hydrostatic equilibrium calculations \citep{Howe2014, Howe2015}, but found that no ring-free scenario was compatible with a stable planetary configuration. Transmission spectroscopy with the HST could not distinguish between the scenarios \citep{Alam2022}. Further atmospheric characterisation with, for instance, the James Webb Space Telescope (JWST) would provide insights into this unusual temperate exoplanet, which differs markedly from the highly irradiated hot Jupiters. Confirming its mass and comparing it with the other planets in the system are crucial steps towards understanding its true nature. The two other long-period planets, d and e, remain incompletely characterised. Only upper limits on the masses of planets d and e were derived in \citet{Santerne2019}, and their orbital periods are still uncertain. With only a single observed transit, HIP 41378 e currently has the least information available. HIP 41378 d was tentatively detected via the Rossiter–McLaughlin effect by \citet{Grouffal2022}, consistent with a period of 278 days. However, a non-detection with CHEOPS reported by \citet{Sulis2024} casts doubt on this estimate, suggesting either that large transit timing variations (TTVs) led to a missed transit or a different orbital period. Based on TESS observations, which never detected a transit of planet d, \citet{Sulis2024} proposed three possible orbital periods: 278, 371, or 1113 days. Determining the true orbital periods and masses of planets d and e is essential for a complete understanding of the HIP 41378 system. An additional non-transiting planet called HIP~41378~g, with a period of nearly 60 days, was proposed as a candidate by \citet{Santerne2019} based on RV analysis. Recently, thanks to TESS and CHEOPS data, \citet{Leonardi2025} measured TTVs longer than three hours for planet c and confirmed the existence of HIP~41378~g.

\begin{figure}[h]
                \centering
                \includegraphics[width=1\columnwidth]{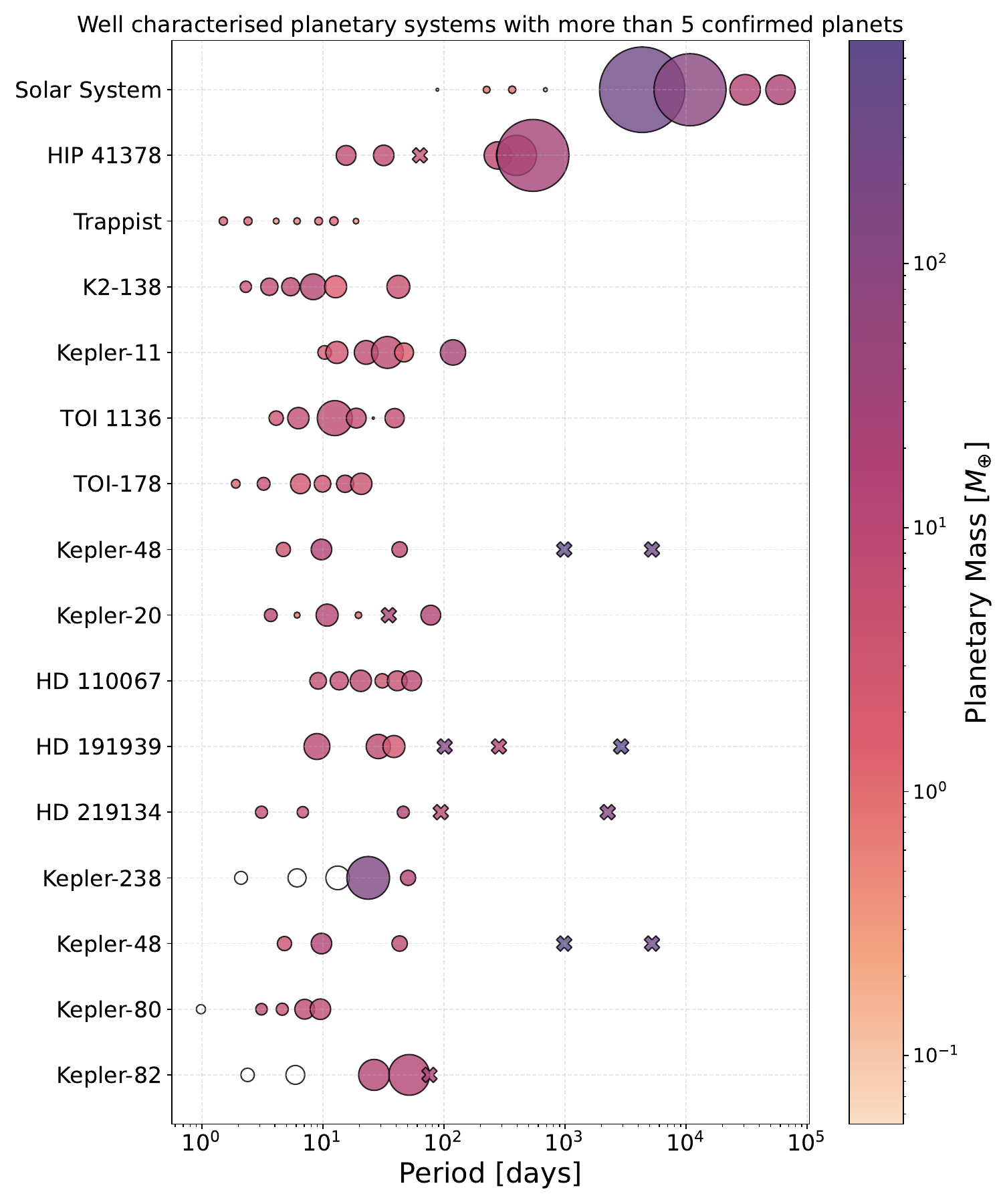}
                   \caption{Confirmed planetary system with more than five planets detected and at least two with determination of both mass and radius. Colours represent planetary masses and circle sizes represent the radius. Planets represented as a cross have no radius measurement, and unfilled circles mark cases where no mass measurement was available. The planetary parameters are from \href{https://exoplanetarchive.ipac.caltech.edu/}{Nasa Exoplanet Archive} }
                          \label{fig_multiplanetary_systems}
\end{figure}

 Figure \ref{fig_multiplanetary_systems} places HIP~41378 in the broader context of confirmed multi-planetary systems with at least five known planets. Only a small fraction of multi-planetary system have precise mass and radius measurements, and very few include long-period giants (P>100 days). HIP 41378 provides a unique opportunity to probe planetary architectures beyond the compact, short-period systems that dominate current populations.

\section{Observations}
\label{sec:obs}
\subsection{Photometry}
\subsubsection{\textit{K2}}

The target HIP~41378 was first observed in photometry during the K2 extended mission \citep{Howell2014} of the Kepler space telescope \citep{Borucki2010}. It was first monitored continuously over $\sim$ 75 days from 27 April 2015 to 10 July 2015 as part of the C5 campaign, in long-cadence mode (30 minutes); it was published in \citet{Vanderburg2016} and re-observed over $\sim$ 52 days from 12 May 2018 to 2 July 2018 as part of the C18 campaign, in short cadence (1 minute) \citep{Berardo2019, Becker2019}.

\subsubsection{\textit{TESS}}
The Transiting Exoplanet Survey Satellite (TESS) observed HIP~41378 during eight sectors: sector 7 ( from January 7 to February 1 2020); sector 34 (from January 14 to February 8 2021); sectors 44, 45, and 46 (from October 12 to December 30 2021); sector 61 (from January 18 to February 12 2023); sector 72 (from November 11 to December 7 2023); and sector 88 (from January 14 to February 11 2025). The data from each sector were processed by the TESS Science Processing Operation Center Pipeline \citep[SPOC, ][]{Jenkins2016}. 

\subsubsection{Other photometric observations}

Transits of HIP41378's planets were also observed by the Spitzer space telescope \citep{Berardo2019}, revealing large TTVs for planet c. These TTVs were further observed by the CHEOPS space telescope and analysed in \citet{Leonardi2025}. The transit of planet f was also observed from the ground by NGTS in 2019 \citep{Bryant2021}; by HST in 2021 \citep{Alam2022}; and by the Tierras, LCOGT, and TRAPPIST-N observatories in 2024 \citep{Garcia-Mejia2025}, also revealing large TTVs for this planet.

\subsection{Radial velocities }

\subsubsection{\textit{HARPS}}

We observed HIP~41378 with the HARPS spectrograph mounted on the ESO 3.6 m telescope at La Silla Observatory in Chile \citep{Pepe2002}. Observations were carried out between 2017-01-27 and 2019-04-18 under ESO programme IDs 198.C-0169 and 0102.C-0171. In total, 370 spectra were obtained over 218 nights, spanning a temporal baseline of 811 days. The median photon-noise uncertainty of the nightly binned RVs is 1.88 m\,s$^{-1}$, with a mean of 1.95 m\,s$^{-1}$, and the RV root-mean-square (rms) amounts to 3.26 m\,s$^{-1}$. These were reduced using the latest data reduction software (v3.2.5), and RVs were binned nightly. A calibration noise of 0.4 \ms\ was added in quadrature to the nightly binned uncertainties.

\subsubsection{\textit{HARPS-N}}

The target HIP~41378 was also observed with the HARPS-N spectrograph \citep{Cosentino2012}, mounted on the \textit{Telescopio Nazionale Galileo} at the Roque de Los Muchachos observatory, in La Palma, Spain. A total of 168 spectra from 2016-02-13 to 2019-05-05 across 161 nights, as part of the GTO programme. The nightly binned RVs have a median photon-noise uncertainty of 1.89 m\,s$^{-1}$, a mean of 2.01 m\,s$^{-1}$, and an RV rms of 3.43 m\,s$^{-1}$. The data were reduced with the latest version of the data reduction software (v3.0.1), and RVs were binned nightly. Similarly to HARPS, a calibration noise of 0.4 \ms\ was added in quadrature to the nightly binned uncertainties.

\subsubsection{\textit{HIRES}}

The HIRES spectrograph collected a total of 75 nightly binned (218 RV measurements in total) epochs between 2016-10-16 and 2019-05-07, spanning a temporal baseline of 942 days. The nightly binned RVs have a median photon-noise uncertainty of 1.19 m\,s$^{-1}$, a mean of 1.36 m\,s$^{-1}$, and an RV rms of 4.81 m\,s$^{-1}$. These observations were conducted using the C2 decker with a typical signal to noise ration (S/N) of 200 per pixel (approximately 250k on the exposure meter, ~5-minute exposures). Wavelength calibration was performed with an iodine cell \citep{Butler1996}. To average over short-timescale stellar oscillations, most HIRES observations were obtained as three consecutive exposures. A high-resolution template spectrum was acquired using the B3 decker on 2016 October 10 under 1.1” seeing. This template consisted of a triple exposure with a combined S/N of 340 per pixel (each exposure reaching 250k on the exposure meter) and was taken without the iodine cell. The HIRES data collection, reduction, and analysis were done according to the California Planet Search methodology \citep{Howard2010}.

\subsubsection{\textit{ESPRESSO}}

We monitored HIP~41378 with the ESPRESSO spectrograph \citep{Pepe2021} mounted on the 8.2-m ESO-VLT telescope at the Cerro Paranal observatory, in Chile.  Observations were carried out between 2020-12-15 and 2024-02-17 as part of programme ID 105.20G0. A total of 272 nightly binned spectra were obtained over 272 nights, spanning a temporal baseline of 1159 days. The nightly binned RVs have a median photon-noise uncertainty of 0.63 m\,s$^{-1}$, a mean of 0.68 m\,s$^{-1}$, and an RV rms of 2.78 m\,s$^{-1}$. The spectra were also reduced with the most up-to-date data-reduction software v3.3.1.

\subsubsection{Other spectroscopic observations}

The host star was also observed by the PFS and the SOPHIE spectrographs as reported in \citet{Santerne2019}. However, only 20 and 11 (respectively) epochs were secured with these instruments over a relatively short time span. Given that HIP~41378 hosts five transiting exoplanets at long orbital periods, as well as extra non-transiting planets and a low-amplitude stellar variability (see Sections \ref{sec:analyse_RV} and \ref{sec:transit_RV}), these time series are not long and dense enough to significantly constrain the system parameters. However, they add complexity in the modelling to account for their potential systematics. As a consequence, we did not include them in the subsequent analysis, but we reported them in this manuscript for completeness and transparency. 

The star was also observed in spectroscopy during the transit of planet d in 2019 and 2022 \citep{Grouffal2022} and planet f in 2022 \citep{Grouffal2025}. These in-transit data were not used in the analysis below.

\section{Radial-velocity analysis}
\label{sec:analyse_RV}

We first analysed the RV data from HARPS, HARPS-N, HIRES, and ESPRESSO independently of the photometry, while adopting constraints from the transiting planets with well-determined orbital periods, namely planets b, c, and f. The RV modelling was performed using the recently developed \texttt{jaxoplanet} package \citep{Hattori2024}, which extends the functionality of \texttt{exoplanet} \citep{exoplanet:joss} and \texttt{starry} \citep{Luger2019} within the JAX framework. JAX enables high-performance numerical computation with automatic differentiation and just-in-time compilation, significantly improving the efficiency of Markov chain Monte Carlo (MCMC) inference.

\subsection{Planet detection}

\begin{figure}[h]
                \centering
                \includegraphics[width=1\columnwidth]{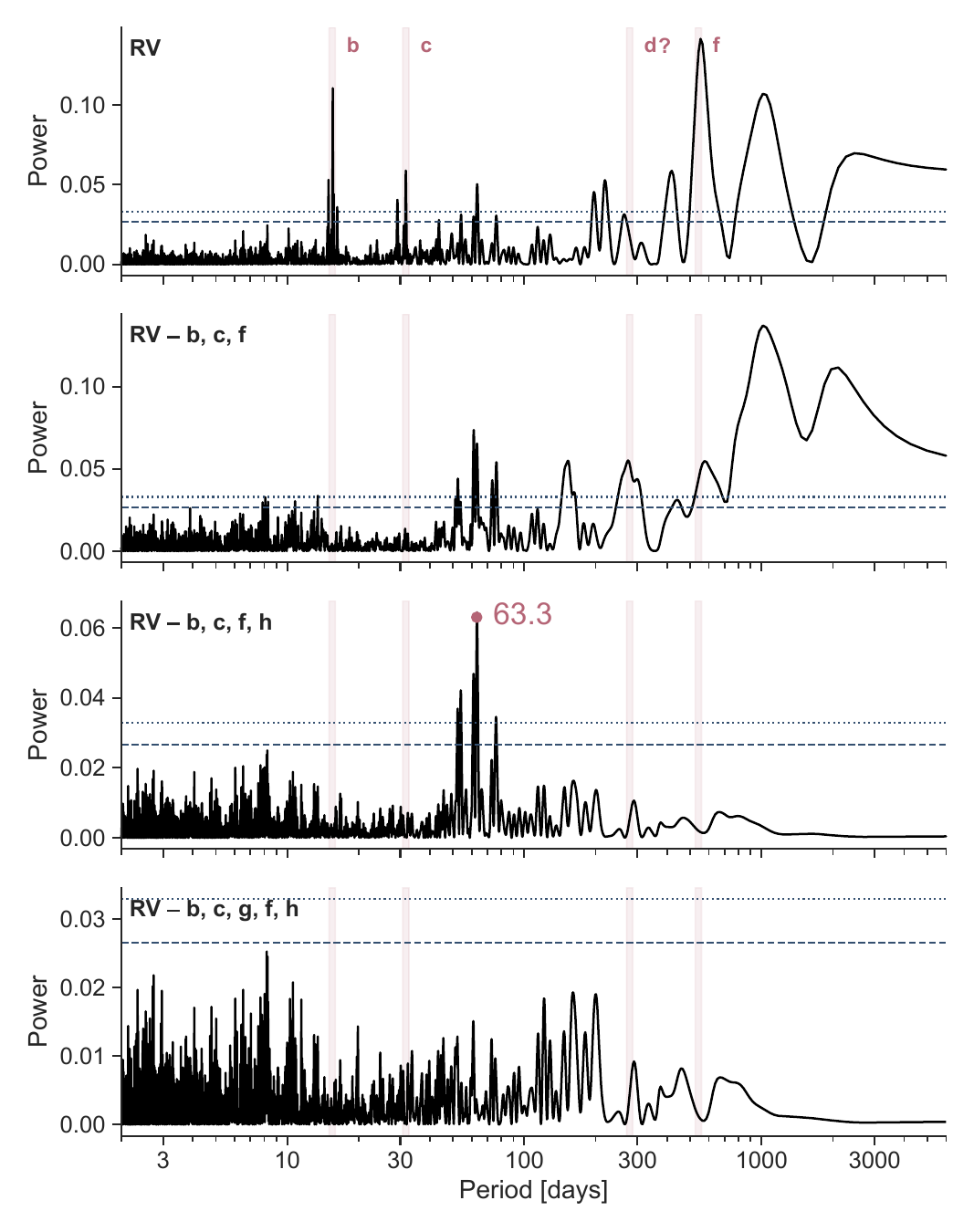}
                   \caption{GLS periodograms of HARPS, HARPS-N, HIRES, and ESPRESSO data. Planet periods b, c, and f are highlighted in pink, as well as the candidate period of planet d (278). Subsequent panels show periodograms after subtracting b, c, and f, and then including h and g. False-alarm probability levels at 1$\%$ and 10$\%$ are indicated as horizontal dotted black lines.}
                          \label{fig_periodograms}
\end{figure}

Preliminary exploration of the RV time series using the generalised Lomb–Scargle (GLS) periodograms \citep{Zechmeister2009} revealed significant signals corresponding to planets b (15.5 days), c (31.7 days), and f (542 days), consistent with their transit ephemerides (upper panel of Fig.~\ref{fig_periodograms}).
We modelled these three planets using the full RV dataset, imposing Gaussian priors on their orbital periods and transit times centred on the observed values. RV semi-amplitudes were assigned uniform priors between 0 and 10~\ms. Eccentricities followed a truncated normal distribution with $\sigma = 0.083$ \citep{VanEylen2019}, and the argument of periastron was uniform over $[-\pi,\pi]$, with the parametrisation in $e\cos\omega$ and $e\sin\omega$ for numerical stability. Instrumental offsets were given broad normal priors centred on the median RV of each dataset, and jitter terms followed half-normal distributions. Posterior sampling was performed with a Hamiltonian Monte Carlo method using the No-U-Turn Sampler (NUTS; \citealt{Duane1987,Hoffman2011}) as implemented in \texttt{NumPyro} \citep{phan2019,Bingham2019}. Four chains were run with 2000 warm-up steps and 3000 draws each (target acceptance 0.99; maximum tree depth 12). 

After subtracting planets b, c, and f, the GLS periodograms show additional signals at 63.3 days and near 278 days, and a long-period variation beyond 1000 days (second panel of Fig.~\ref{fig_periodograms}). The latter was modelled with a Keplerian with a uniform prior between 800 and 4000 days, converging to a candidate solution at $2828 \pm 287$ days (planet candidate h). Including planets b, c, f, and h, the 63-day signal remains highly significant and matches the candidate planet g reported by \citet{Santerne2019} ($P = 62.06 \pm 0.32$ days).

With planets b, c, f, g, and h included, no significant peaks remain in the periodogram at the expected periods of planet d (bottom panel of Fig.~\ref{fig_periodograms}), and no additional signals are detected. However, the existence of planets d and e is proven by their transits, implying RV semi-amplitudes below the current detection threshold and suggesting that both planets are low mass.

\subsection{Model comparison for planets d and e}
\label{sec:model_comparison}

The orbital periods of HIP~41378~d and e cannot be constrained from photometry alone, and no significant signals are detected in the periodograms that would allow an unambiguous identification in the RV data. Planet d has three candidate periods: 278, 371, and 1113 days \citep{Sulis2024}, the longest of which requires a high eccentricity to reproduce the observed transit. To investigate which of these configurations is most consistent with the RV data, we performed a series of fits and model-comparison tests, progressively increasing the model complexity.

\subsubsection{Markov chain Monte Carlo analysis}
\label{sec:RV}
\begin{figure}[h]
		\centering
		\includegraphics[width=1.0\columnwidth]{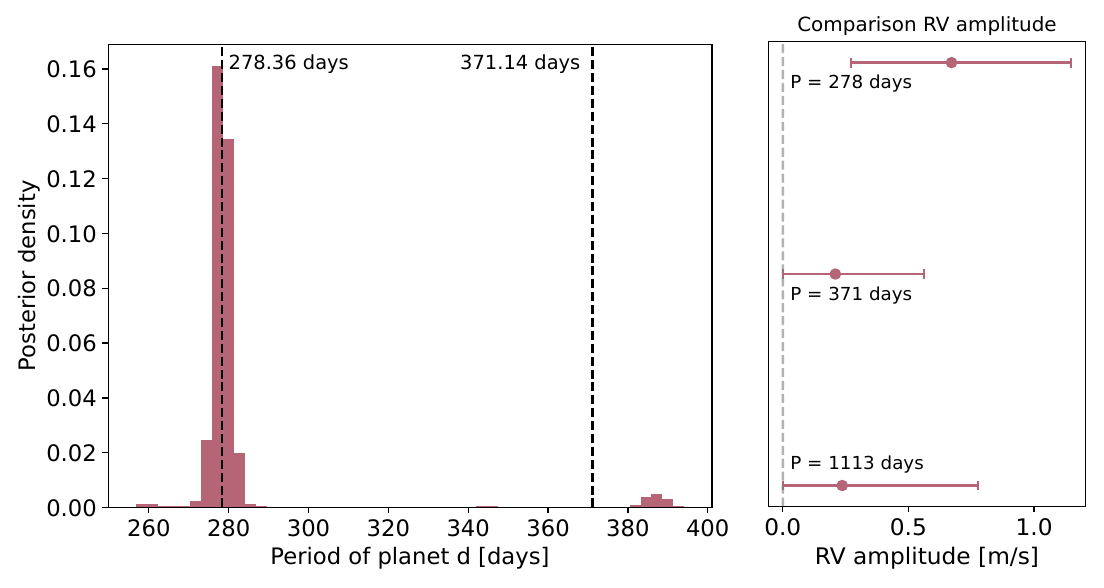}
		   \caption{Tests for the orbital period of HIP~41378~d. \textit{Left panel:} Posterior density of the period of planet d, considering a uniform prior between 250 and 400 days. The vertical dashed lines indicate the possible periods from transit observations. \textit{Right panel:} Resulting RV semi-amplitude and 99$\%$ HDI for the three periods tested: 278, 371, and 1113 days.}
			  \label{fig_model_comp}
\end{figure}

We first modelled the full RV dataset, including all confirmed planets except planet e, to isolate the contribution of planet d. A broad uniform prior was adopted for the period of planet~d between 250 and 400~days, encompassing the two most plausible solutions (278 and 371~days). The resulting posterior distribution (left panel of Fig.~\ref{fig_model_comp}) is strongly concentrated near $P_d \simeq 278$ days, while secondary modes at longer periods (e.g. $\sim$385 days) have significantly lower probability densities. This preliminary analysis disfavours the 371-day solution.

However, the 278-day period is inconsistent with the non-detection of a transit by CHEOPS in 2023 \citep{Sulis2024}. We therefore explored three discrete orbital periods for planet d (278, 371, and 1113 days), again excluding planet e. The right panel of Fig.~\ref{fig_model_comp} shows the median and 99.5$\%$ highest density interval (HDI) for the RV semi-amplitude of planet d under these assumptions. Only the 278-day model yields a significant ($>3\sigma$) non-zero semi-amplitude.

To assess whether the preference for the 278-day signal could instead be attributed to planet e, we extended the analysis by explicitly including planet e and performing model comparison using leave-one-out cross-validation (LOO CV). This approach provides a robust estimate of out-of-sample predictive performance \citep{Vehtari2015, Vehtari2024} and is well suited to high-dimensional RV models. The LOO estimates were computed from the posterior samples obtained with \texttt{NumPyro}---without rerunning the MCMC for each data subset---using Pareto-smoothed importance sampling \citep{Vehtari2015} provided in the Python package \texttt{ArviZ}.

\begin{figure}[h]
                \centering
                \includegraphics[width=1\columnwidth]{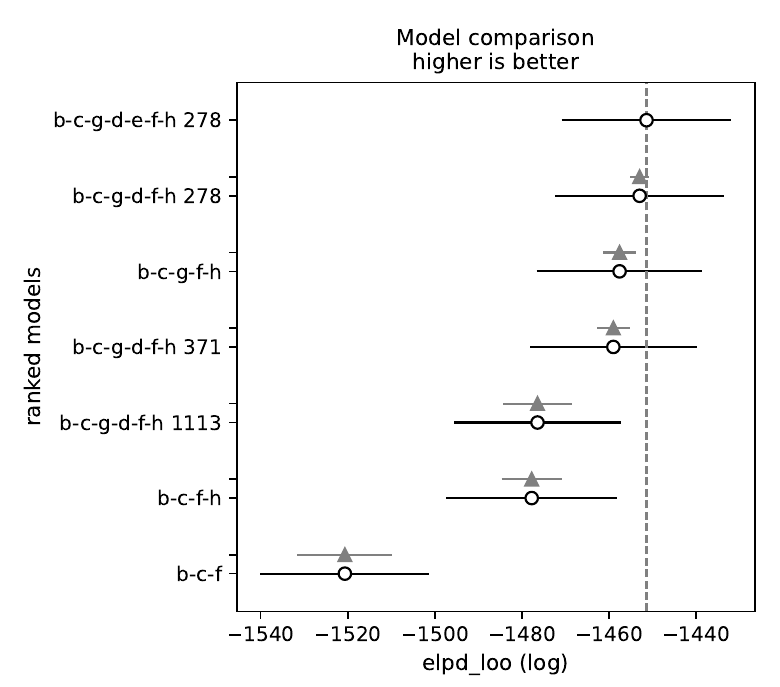}
                   \caption{Comparison of different RV models, including subsets of the HIP~41378 planets, ranked by their expected log point-wise predictive density (elpd LOO). Each model label indicates the included planets and, for planet d, the tested orbital period in days. Higher elpd LOO values correspond to better out-of-sample predictive performance. The horizontal line segments represent the standard error of the elpd. The lighter line segment with triangles indicates the elpd difference of one model relative to the best model.}
                          \label{fig_comparison_models}
\end{figure}

For each assumed value of $P_d$, we explored different uniform period ranges for $P_e$, reflecting the possible relative orbital ordering of the two planets:  (1) if $P_d = 278$~days, we tested three possible ranges for $P_e$: first, assuming planet e lies interior to planet d (100–278 days), second, assuming planet e lies between planets d and f  (300 - 540 days), and third, assuming planet e lies outside planet f (600- 1000~days); (2) if $P_d = 371$~days, we tested two ranges: 100–370 and 600–1000~days; (3) if $P_d = 1113$~days, we adopted a broad range: 100–540~days. Only models with well-converged posteriors (i.e. $\hat{R} \approx 1$ and minimal divergences) were included. Here, $\hat{R}$ denotes the Gelman–Rubin statistic, which measures MCMC convergence by comparing within and between-chain variances. The results, ranked by expected log point-wise predictive density (elpd LOO), are shown in Fig.~\ref{fig_comparison_models}.

The best-performing model includes all seven planets with $P_d = 278$ days and $P_e$ between 300 and 540 days. Models without planet e or with $P_d = 371$ days remain statistically competitive, but they generally favour a signal near $P_e \sim 278$ days, which strongly suggests the presence of a planet at this period.

\subsubsection{Nested sampling}

Given the low RV semi-amplitude and the limited transit constraints for planets d and e, we performed an independent nested sampling analysis using RV-only data first, with HARPS, HARPS-N, and ESPRESSO data. Orbital periods for planets b, c, and f were constrained with Gaussian priors, while broad uniform priors were adopted for planets g ($P_g \in [61,65]$ days) and the long-period candidate h ($P_h \in [800,4000]$ days), allowing us to focus on the poorly constrained periods of planets d and e.

We used \texttt{jaxns} \citep{Albert2020} with 900 live points. For planet d, we tested the two possible periods of 278 and 371 days using a uniform prior between 200 and 400. The nested sampling analysis strongly favours the 278-day solution.

For planet e, we adopted a broad uniform prior ($P_e \in [150,450]$ days). The posterior distribution is dominated by the $\approx390$ day solution, with less significant secondary peaks at shorter periods (see Fig. \ref{fig_nested_sampling}).

To better constrain the orbital period of planet e, we performed three tests combining photometric and RV data, considering configurations in which planet e is located inside planet d, between planets d and f, or outside of planet f. The configuration in which planet e lies between planets d and f is strongly favoured, with a difference in Bayesian evidence of $\Delta \log Z \simeq 15$ compared to the scenario in which planet e is exterior to planet f, and $\Delta \log Z \simeq 6.5$ compared to the configuration in which planet e is interior to planet d.
These results are consistent with the MCMC analysis and model comparison with LOO-CV. The complete nested sampling analysis can be found in Appendix \ref{appendice:nested_sampling}.

\subsection{RV analysis and instrument-specific datasets}

\subsubsection{Global fitting}

For the global data fitting, we adopted Gaussian priors on the orbital periods of planets b, c, and f, centred on the transit-derived values with narrow standard deviations. Uniform priors were used for the remaining planets: 60–65 days for g, 250–400 days for d, 360–450 days for e, and 700–4000 days for h. These priors are based on two assumptions based on knowledge from photometry and model comparisons from Section \ref{sec:model_comparison}: the period of planet d is either 278 or 371 days, and we consider that planet e is before planet f in the system, considering its shorter transit duration ($\approx$ 13 hours) and assuming low eccentricities in the multi-planetary system. The analysis was carried out sequentially, first using HARPS, HARPS-N, and HIRES data as in \citet{Santerne2019}, then ESPRESSO data alone, and finally the full dataset combining all four instruments.
Again, we used \texttt{NumPyro} to run four MCMC chains in parallel, each with 2000 warm-up steps and 3000 draws.
The posteriors for periods and RV semi-amplitudes are shown in Fig.~\ref{fig_posterior_priors_RV}. The RV signal of planet e becomes clearly detectable only when ESPRESSO data are included, reflecting its low semi-amplitude and incomplete phase coverage in the other datasets. The period of planet d consistently converges to 278 days across all instruments. The candidate planet h shows the largest variation between instruments, being most visible in ESPRESSO data. 

The preferred model, with all planets included, still shows some correlated residuals, with an rms of 2.60~\ms. The HIRES data (with 75 measurements over a total of 713) show the largest errors and the largest residual rms, with a weighted rms of 4.22~\ms. Excluding the HIRES data improves the overall fit, reducing the weighted rms of the residuals to 2.38~\ms. Nevertheless, some correlations persist and are likely related to stellar activity. We therefore investigate the impact of stellar activity in Section \ref{sec:stellar_act}.

\subsubsection{Blind search for RV signals}
\label{sec:l1}

We performed a fully blind search for periodic signals using the $\ell_1$ periodogram \citep{Hara2017}, which reconstructs the RV time series as a sparse sum of sinusoids on a predefined frequency grid. The method takes the RV data as input, a frequency grid, a noise covariance model, and a linear base model, and returns a periodogram-like output with reduced aliasing and associated FAP estimates.

\begin{figure}[h]
                \centering
                \includegraphics[width=1\columnwidth]{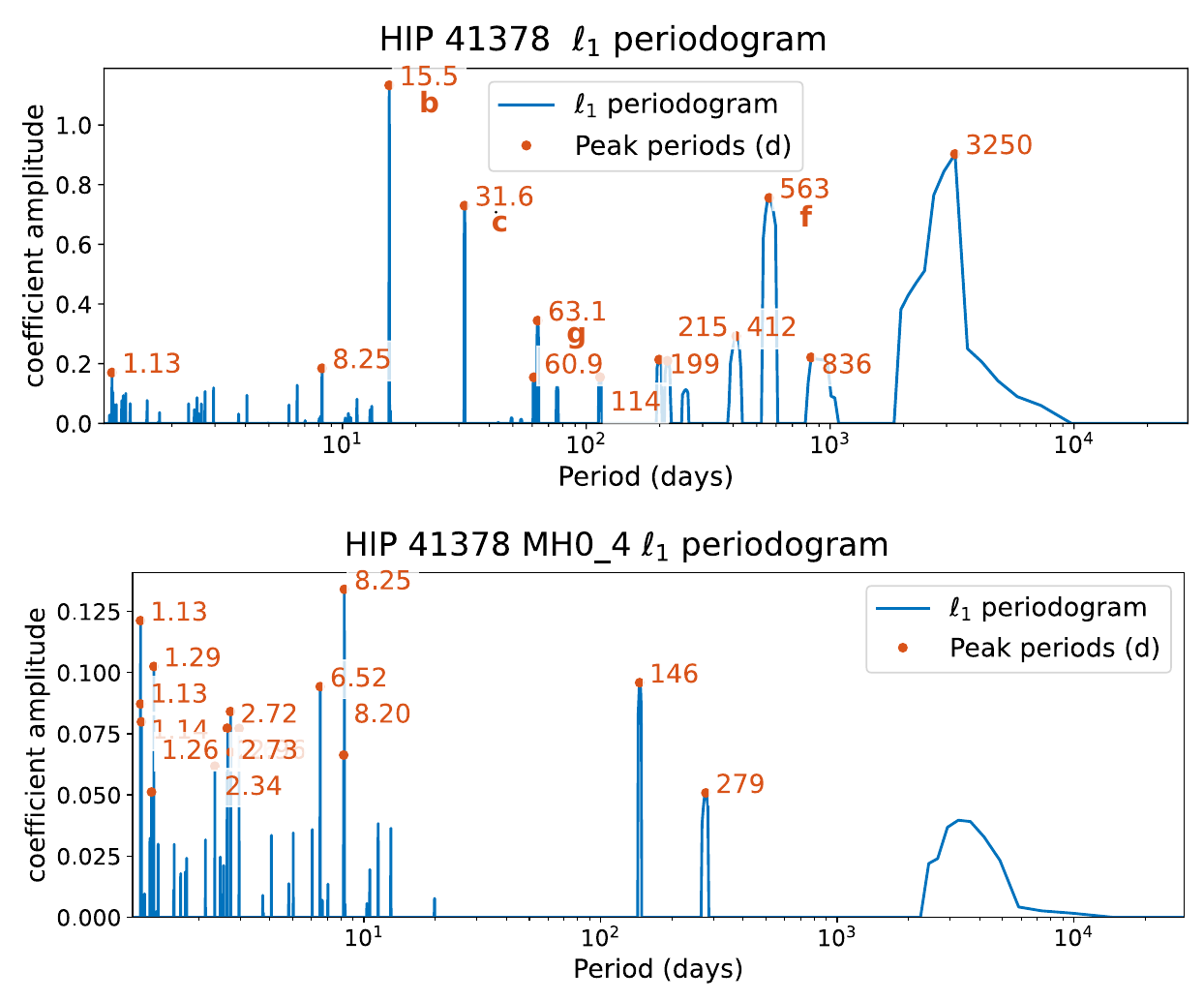}
                   \caption{\textit{Upper panel}: $\ell_1$ periodogram of the RV data, including offsets in the base model. \textit{Lower panel}: $\ell_1$ periodogram of the RV data, including offsets, transiting planets, and order 11 polynomial in the base model. Planets b, c, g, and f are highlighted by their letters.}
                          \label{fig:l1_full}
\end{figure}

We analysed the combined HIRES, HARPS, HARPS-N, and ESPRESSO data over a frequency range of 0–0.95 cycles/day with a step of $1/(10T_\mathrm{obs})$. The noise was assumed white, with instrument-specific jitter terms fixed to the fitted values (Table~\ref{posteriors}). As a linear base model, we first only used the offsets of each instrument. The results are shown in Fig.~\ref{fig:l1_full}; the known planets are clearly visible, as is a long-period signal at 3250 d. The $\log_{10}$ FAPs of the 15, 31, 63, 562, and 3250 d signal are -15.5, -14.8, -5.2, -22.1, and -15.9. We note that the only other significant peak is at 835 d, with a $\log_{10}$ FAP of -3.6; however, as discussed below, the interpretation of this signal is difficult.

We next refined the analysis in two ways. First, we note that the $\ell_1$ periodogram seeks to find a Fourier spectrum that reproduces the data but has as many vanishing values as possible. This was done by minimising the sum of the absolute values of the Fourier amplitudes, that is, the $\ell_1$ norm of the Fourier spectrum. As a consequence, amplitudes were underestimated, which might have impacted our ability to detect the smallest signals. To circumvent this issue, following \citep{hara2020}, we included sinusoids at the periods of the confirmed transiting planets (15.57209, 31.70591, 63.19, 542.07975 d) in the base model. The asymmetric 3250 d peak suggests that a purely periodic model is inadequate. We therefore added a low-amplitude polynomial trend. Testing polynomial degrees from 5 to 15 and selecting via cross-validation \citep{hara2020}, we adopted degree 11 (lower panel of Fig.~\ref{fig:l1_full}). In this configuration, no additional signals reach formal significance (FAP > 5\%). However, there are signals at 6.5 and 8 days (1.13 d is an alias of 8.48 d), at the expected stellar rotation period (see Section \ref{sec:stellar_act}). In Appendix~\ref{appendice:asp}, we show that this signal is not stable over time, which supports the hypothesis that it is of stellar origin.  

Additionally, the $\ell_1$ periodogram exhibits a peak at 279 d, which is fully compatible with the 278 d period of planet d. We finally note a 146 d signal. A planet at this period would continue the resonance chain, and could be due to a low-mass, non-transiting planet (see Section \ref{sec:resonances}). Although the 279 signal is not statistically significant, finding it at this particular period, without any prior on the position of putative signals, is a strong hint of its planetary origin. 

\subsection{Stellar activity analysis}
\label{sec:stellar_act}
\subsubsection{Stellar rotation period}
The HIP~41378 object is a relatively quiet F-type star. An analysis of the K2 short-cadence photometry by \citet{Grouffal2025} yielded a stellar rotation period of P$_{rot}$ = 6.8 $\pm$ 0.4 days, which was derived using a quasi-periodic Gaussian process (GP) model of the light curve.

The average Ca II chromospheric activity index $\log(R'_{HK})$ is measured as -4.88 $\pm$ 0.03 for ESPRESSO, -4.94 $\pm$ 0.02 for HARPS, and -4.95 $\pm$ 0.02 for HARPS-N, indicating a low and stable level of stellar activity over time. Using the empirical relations from \citet{Noyes1984} and \citet{Mamajek2008}, these values translate into an estimated stellar rotation period of 7.75 $\pm$ 1.42 days according to ESPRESSO and 8.29 $\pm$ 1.52 days for HARPS and HARPS-N. These estimates are consistent with the photometric rotation period within uncertainties.

\subsubsection{Frequency analysis of the activity indicators}
\label{sec:frequency_activity_indicators}

We investigated several stellar activity indicators using HARPS, HARPS-N, and ESPRESSO data, including the full width at half maximum (FWHM) of the cross-correlation function (CCF), the bisector span (BIS; \citealt{Queloz2001}), the CCF contrast, and absorption-line diagnostics such as the Ca~II H$\&$K S-index and the H$\alpha$ line. A detailed description of this analysis is provided in Appendix~\ref{sec:stellar_activity}. Of particular interest is the long-period RV signal attributed to candidate planet h. If this signal were present in the activity indicators, it could point to a stellar origin, such as a magnetic cycle. All indicators except the BIS exhibit significant long-period power (periods $\gtrsim$ 900 days), suggesting the presence of long-term stellar activity. However, the orbital period of the candidate planet is longer than the signal observed in the activity indicators ($\sim$ 1000--1700 days). 

After subtracting the long-period component from the FWHM, contrast, and S-index using sinusoidal fits, no significant power remains at shorter periods associated with stellar rotation. The BIS only shows a very low-amplitude signal at $\sim$ 8 days, but a quasi-periodic GP fit does not converge. Overall, no robust rotational signal is detected in any activity indicator.

In summary, HIP~41378 exhibits evidence of long-period stellar activity variability, likely related to a magnetic cycle, which may partially contribute to the RV signal. In contrast, the absence of a clear rotational signature in the activity indicators confirms the star’s low activity level. The star’s inclination of $i_* = 42^{+6}_{-7}$~deg measured through the Rossiter--McLaughlin effect and considering the rotation period \citep{Grouffal2025} may suppress the observed rotational modulation, consistent with the low amplitude in both photometry and activity indicators.

\subsection{1D GP analysis}

After modelling the six confirmed planets and the long-period signal of the candidate planet h, correlated residuals remain in the RVs. The periodogram of the residuals (Fig. \ref{fig_residuals}) reveals excess power during short periods of $\sim$8 days, similar to the stellar rotation period, suggesting a stellar activity origin. To account for this correlated noise, we added the fit of a GP to the fit of the seven planets in the RV data of the four instruments. GPs are a popular mathematical tool for modelling stellar activity-induced signals that describe stochastic variations (see \citealp{Aigrain2023} for a complete review).

We adopted a quasi-periodic (QP) kernel of the form

\begin{equation}
\label{eq:QP_kernel}
k(t, t') = A^2 
\exp\!\left[-\frac{(t - t')^2}{2\,\ell^2}\right]
\exp\!\left[-\Gamma \, \sin^2\!\left(\frac{\pi (t - t')}{P_{\mathrm{rot}}}\right)\right],
\end{equation}where $A$ is the GP amplitude, $\ell$ is the exponential decay timescale, $P_{\mathrm{rot}}$ is the QP timescale associated with stellar rotation, and $\Gamma$ controls the coherence of the periodic component.
The GP kernel was implemented with \texttt{tinyGP} \citep{Foreman-Mackey2024}. We adopted uniform priors on all the GP hyperparameters: $P_{\rm rot} \in [5,10]$ days, $A \in [0.01,5.0]$~\ms, $\ell \in [1,500]$ days, and $\Gamma \in [0.1,10]$. Posterior sampling was performed using four parallel chains with 2000 warm-up steps and 3000 draws each. The planetary system parameters were given the same priors as in the RV-only analysis. The fit converged to a stellar rotation period of $8.12 \pm 0.09$~days, consistent with the $\log(R'_\mathrm{HK})$ estimate and the signal observed in the residuals' periodograms. The characteristic evolution timescale of active regions converged towards $l = 100_{-70}^{+60}$, indicating long-lived surface structures persisting over $\approx$ 10–15 stellar rotations and compatible with a star with low activity. The $\Gamma$ parameter converged towards $\Gamma = 0.94_{-0.83}^{+0.52},$ which could indicate the signature of a small number of dominant active regions.

We note a modest discrepancy between the photometric rotation period ($\sim$6–7 days) and the RV-derived period ($\sim$8 days). This difference may be explained by a combination of stellar inclination, differential rotation \citep{Grouffal2025}, and the latitude distribution of active regions. Photometric variability is dominated by equatorial spots, which rotate faster under differential rotation, whereas RVs can be more sensitive to higher latitude structures that may remain visible throughout the stellar rotation.

\begin{figure}[h]
                \centering
                \includegraphics[width=1\columnwidth]{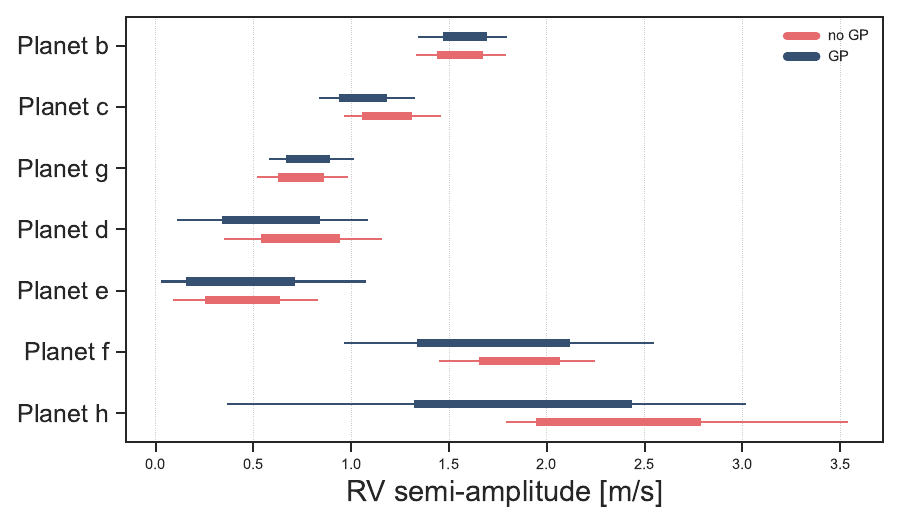}
                   \caption{Comparison of the semi-amplitude of the seven planets in the HIP~41378 system for a model with GPs (in blue) and without GPs (in pink). The 68$\%$ HDI is represented as a box and the 99$\%$ HDI as a horizontal bar.}
                          \label{fig_comparison_K_GP}
\end{figure}

Including the GP in addition to the planetary system parameters significantly improves the RV fit. The rms of the residuals decreases from 2.38 to 2.16~\ms, and the GP model is strongly favoured by LOO CV, with $\Delta \mathrm{elpd}_{\mathrm{loo}} \approx +24$ relative to the no-GP model. Figure~\ref{fig_comparison_K_GP} compares the resulting posterior distributions of the RV semi-amplitudes.

Including a GP increases uncertainties on the semi-amplitudes of the long-period planets, as expected. The GP absorbs part of the long-timescale stellar variability, preventing the orbital model from artificially fitting activity-induced trends. This leads to more conservative and realistic estimates of the RV semi-amplitudes: $K$.
For these reasons, we adopted the GP model as the final RV model for subsequent analyses combined with photometric data.

We also tested a multi-dimensional GP framework \citep{Rajpaul2015} incorporating activity indicators and their derivatives. However, given the absence of a clear rotational signal in the activity indicators, this approach did not yield a significant improvement over the 1D GP model.

\section{Joint analysis of photometry and RV}
\label{sec:transit_RV}

\begin{figure*}[h]
   \centering
   \includegraphics[width=1.\textwidth]{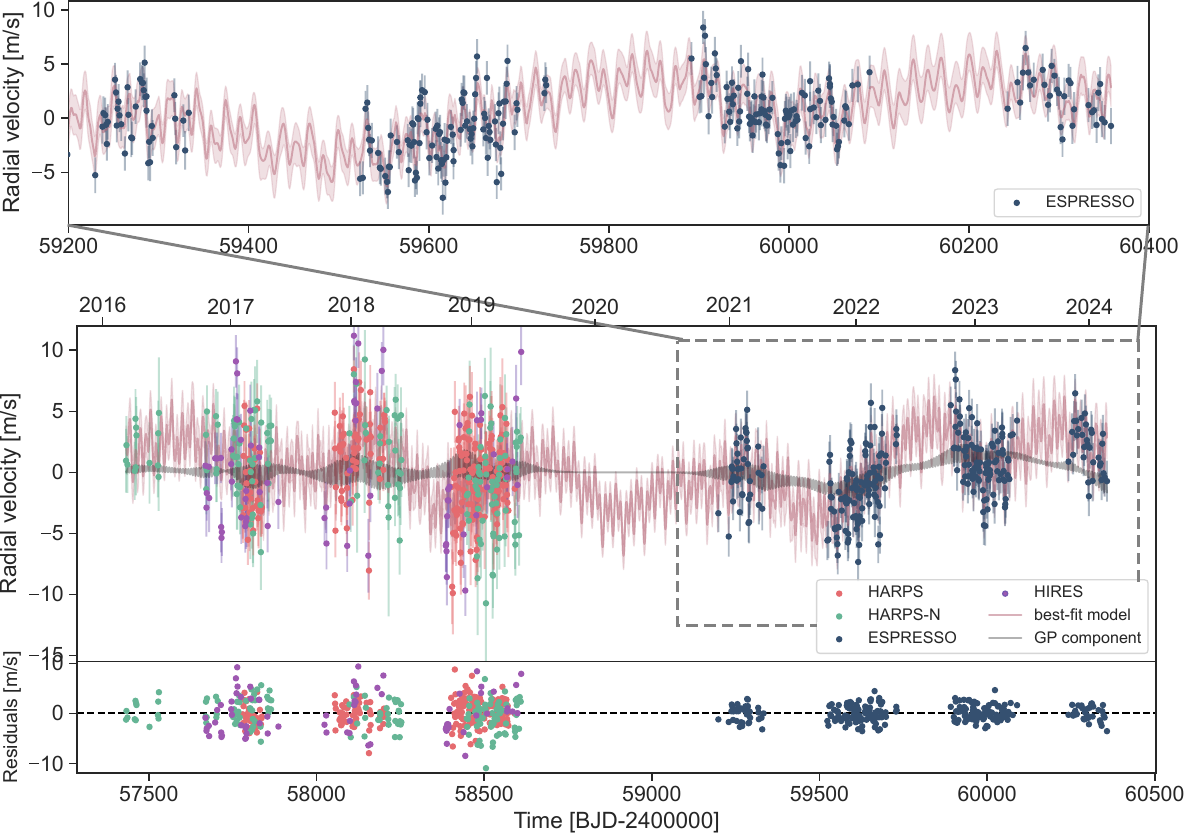}
      \caption{RV timeseries of HIP~41378 with HARPS (pink), HARPS-N (green), HIRES (purple) and ESPRESSO (blue), along with best fit model (median of the posterior) in pink and the GP component in black. The upper panel is a zoom on ESPRESSO data. The lower panel represents the residuals of the fit (data - model).}
         \label{fig_final_fit}
   \end{figure*} 

To jointly constrain the orbital and physical parameters of the HIP~41378 planetary system, we performed a combined analysis of K2 photometry and RV data from HARPS, HARPS-N, HIRES, and ESPRESSO using \texttt{jaxoplanet} \citep{Hattori2024}. Only K2 light curves were included, as no transits of planets d, e, or f have been detected with TESS or CHEOPS. Independent fits of TESS and CHEOPS data for planets b and c, including transit-timing variations of planet c, are presented in \citet{Leonardi2025}. Including these datasets did not improve constraints on the outer planets.

The priors are listed in Table~\ref{posteriors}. They assume the best model found with model comparison from the RV-only analysis.

For transiting planets, we imposed informative priors on the orbital inclination and sampled uniformly in $\cos{i}$. Planetary masses were derived from the inclination-dependent RV semi-amplitudes.
Transit depths were parametrised in log-space with Gaussian priors centred on the values from \citet{Vanderburg2016}, and planetary radii were computed from these depths. Bulk densities were derived from the sampled mass and radius. Quadratic limb-darkening coefficients were sampled using the parametrisation from \citet{Kipping2013}.

For non-transiting planets, the time of inferior conjunction, $T_0,$ was not directly observed. We therefore parametrised the orbital phase using a dimensionless parameter, $\phi$, which represents the phase offset relative to the mean observation time. The reference epoch was then defined as $t_0 = t_{ref} + \phi \times P,$ where  P is the orbital period and $t_{ref}$ is the mean of the observation times. This parametrisation centres the epoch near the dataset, reducing parameter degeneracies and numerical correlations during sampling.

The sampling was performed using NUTS \citep{Duane1987,Hoffman2011} as implemented in NumPyro \citep{phan2019, Bingham2019}. Three chains were run in parallel, each with 2000 warm-up steps and 3000 draws, with a target acceptance probability of 0.99. Using NUTS in JAX allows for fast gradient-based sampling, even for high-dimensional problems; the full RV analysis with four instruments and 43 free parameters completes in minutes. Achieving stable convergence required non-centred re-parametrisation of latent variables. In hierarchical models, rather than sampling directly from $\theta \sim \mathcal{N}(\mu, \sigma)$, we sampled $z \sim \mathcal{N}(0,1)$ and set $\theta = \mu + \sigma \cdot z$. This avoided funnel-shaped posterior geometries and improved sampler efficiency. This approach was applied to both Gaussian and uniform priors. A quasi-periodic GP was included in the RV model to account for stellar activity using the same priors as the 1D-GP analysis.

Figure  \ref{fig_final_fit} shows the RV time series with the median posterior model, the GP component, and residuals. Phase-folded RVs for each planet are shown in Figure~\ref{fig_phase}. The main results of the joint fit are summarised in Table~\ref{tab_parameters}, with full posteriors and priors in Table~\ref{posteriors}. Figure \ref{fig_fit_transit} shows the two K2 light curves with the median of the best-fit model.

All planetary masses are below 10\Mearth, except for planets f and h. We derived $m_b = 6.9 \pm 0.5$\Mearth, $m_c = 5.97^{+0.56}_{-0.86}$ \Mearth, $m_g = 5.75^{+0.76}_{-0.83}$ \Mearth, $m_d = 6.53^{+2.65}_{-3.15}$ \Mearth, $m_e = 7.62^{+3.20}_{-4.63}$ \Mearth, $m_f = 25 \pm 5$ \Mearth, and $m_h = 43^{+16}_{-13}$ \Mearth. For planets g and h, the reported values are minimum masses due to the unknown inclination. The mass of planet f is significantly higher than the $12 \pm 3$\Mearth\ previously reported by \citet{Santerne2019}, likely due to the inclusion of ESPRESSO data and improved modelling of long-period signals. These mass estimates lead to the following bulk densities for the five transiting planets: $\rho_b = 2.07 \pm 0.16$\gcm, $\rho_c = 1.60 \pm 0.21$\gcm, $\rho_d = 0.74 \pm 0.33$\gcm, $\rho_e = 0.30^{+0.13}_{-0.18}$\gcm, and $\rho_f = 0.16^{+0.03}_{-0.04}$\gcm. Densities decrease with increasing orbital distance. For reference, Jupiter and Saturn have densities of 1.33 and 0.68~g\,cm$^{-3}$, respectively. Planets e and f are notably less dense than Saturn. All planets exhibit low eccentricities, except candidate planet h. The uncertainty on the orbital period of the candidate planet h remains large, despite its high RV amplitude; further observations are needed to confirm its planetary nature.

\begin{figure}[h!]
                \centering
                \includegraphics[width=1.0\columnwidth]{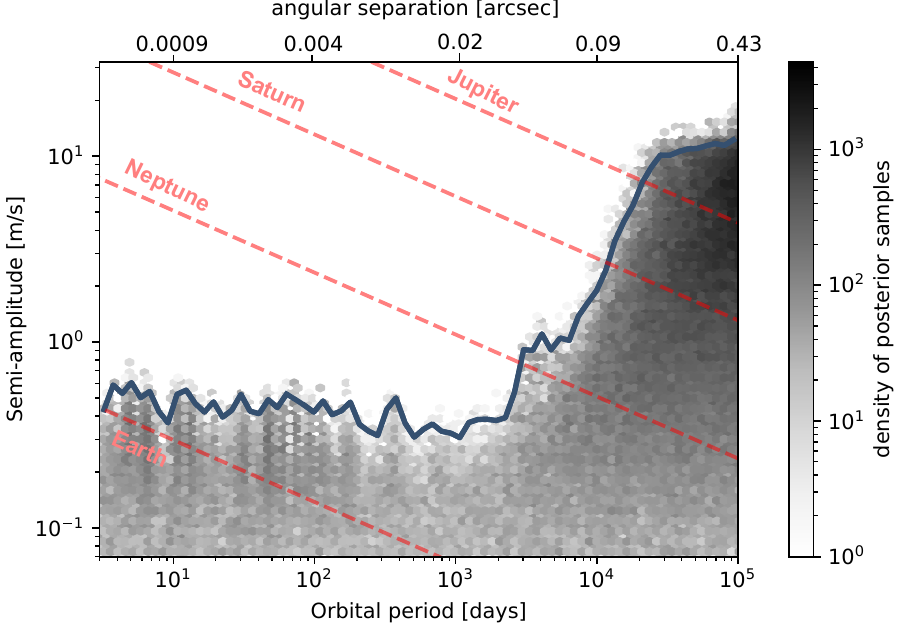}
                   \caption{Greyscale hex-bin plot representing the posterior samples obtained from \texttt{kima} on the full RV dataset residuals with the number of planets fixed to one. The blue line represents the 99th percentile detection limit. The RV semi-amplitudes of Jupiter, Saturn, Neptune, and Earth are represented by dashed pink lines for comparison. The top axis shows the corresponding angular separation.}
                          \label{fig_detection_limit}
\end{figure}

\begin{figure}[h]
                \centering
                \includegraphics[width=1\columnwidth]{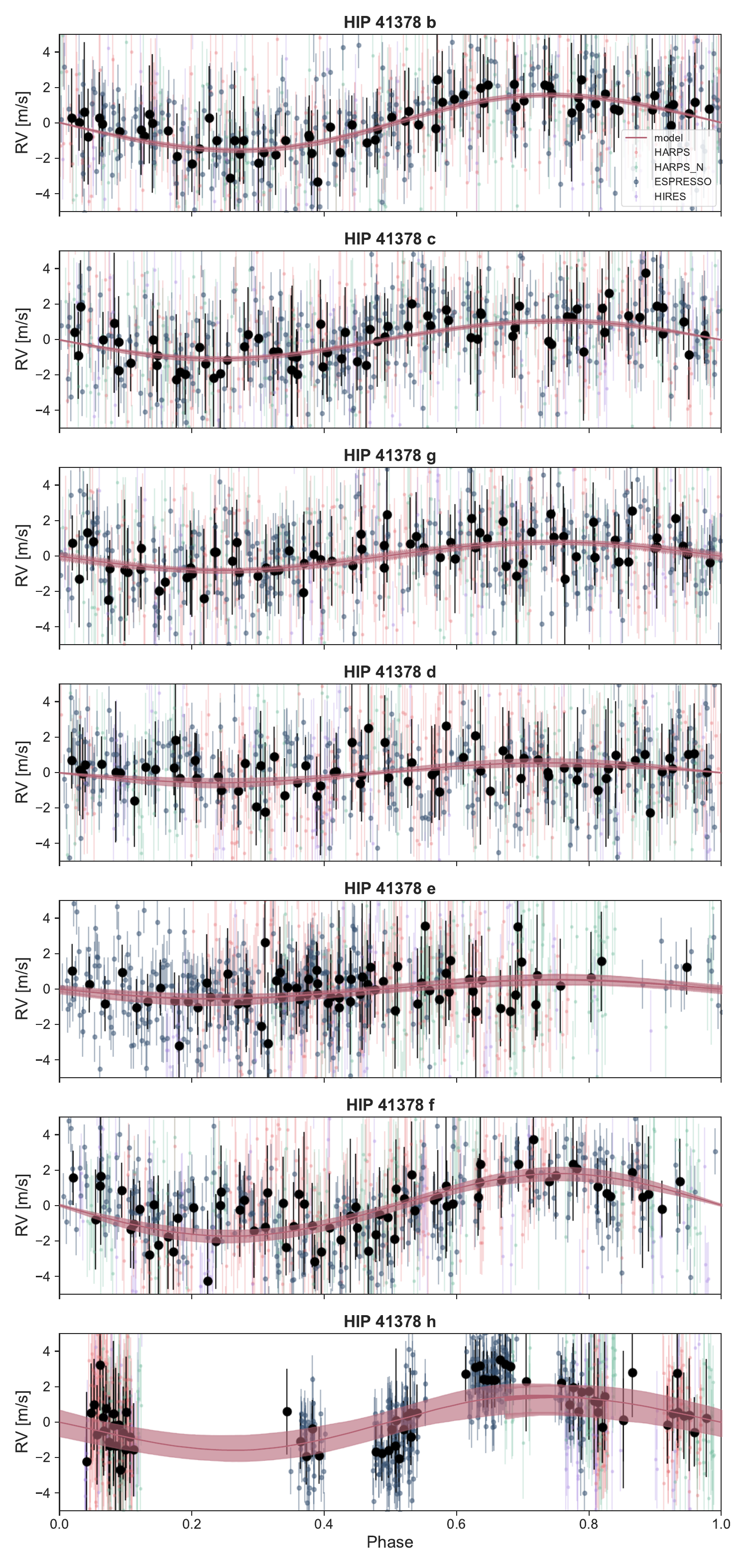}
                   \caption{HARPS-N, HARPS, HIRES, and ESPRESSO RV phase-folded on the results of the joint RV+transit fit for the seven planets in the system. The pink line represents the median of the best-fit model RV curve with its 68$\%$ HDI in the shaded area.}
                          \label{fig_phase}
\end{figure}

We assessed the completeness of our RV analysis by computing the detection limit on the residuals after subtracting the median best fit of planets b, c, g, d, e, f, and the candidate h, as well as the quasi-periodic GP. We used the diffusive nested sampler \texttt{kima} \citep{Faria2018} following the approach of \citet{Standing2022}, \citet{Standing2023}, \citet{John2023}, and \citet{Balsalobre-Ruza2025}. This method maps all the Keplerian signals compatible with the residuals and defines a detection threshold, which is shown in Figure \ref{fig_detection_limit}. Signals lying above this threshold would have been detected in our analysis. We find that no additional planets with orbital periods shorter than $\sim$7 years and RV semi-amplitudes larger than 1~\ms\ can be present in the system. Earth-mass planets remain below the sensitivity threshold of our dataset, while long-period companions such as Jupiter- or Saturn-mass planets at periods $\gtrsim$ 30 years cannot be ruled out. Complementary constraints are provided by astrometric acceleration measurements. HIP~41378 appears in the catalogues of \citet{Kervella2022} and \citet{Brandt2021}. No statistically significant proper-motion anomaly is reported. In the separation range of 3–10 au, where the technique reaches its maximum sensitivity, companions more massive than $\sim$2 Jupiter masses can be excluded. These astrometric limits therefore reinforce our RV detection limits at intermediate separations. The details of the method are provided in Appendix \ref{appendice:detection_limit}.

\begin{table}
\caption{Parameters and posteriors for the transit and RV model.}
\begin{tabular}{lc}
\hline
Parameter  &   Posterior               \\
Planet b  &                    \\
Orbital period P$_b$ [days]   & $15.57209 \pm 0.00002$       \\
Mid-transit time T$_{0,b}$ [BJD]$^a$     & $7152.282 \pm 0.0019$   \\
Orbital eccentricity $e_b$       & $0.05^{+0.03}_{-0.05}$  \\
Planet mass M$_b$ [\Mearth]     & $6.9 \pm 0.5 $   \\
Planet radius $R_b$ [\Rearth]   &  $2.64^{+0.022}_{-0.024}$ \\
Bulk density $\rho_b$ [\gcm] & $2.07 \pm 0.16$\\
Impact parameter b$_b$             & $ 0.448 \pm 0.029$ \\
Transit duration T$_{14,b}$ [h]   & $5.26  \pm 0.21$ \\
\hline
Planet c  &                 \\
Orbital period P$_c$ [days]  & $31.70591 \pm 0.00005$       \\
Mid-transit time T$_{0,c}$ [BJD]$^a$   &   $7163.166^{+0.0016}_{-0.0013}$        \\
Orbital eccentricity $e_c$         & $0.044^{+0.022}_{-0.044}$  \\
Planet mass M$_c$ [\Mearth]     & $5.97^{+0.56}_{-0.86}$   \\
Planet radius $R_c$ [\Rearth] & $2.74^{+0.043}_{-0.049}$ \\
Bulk density $\rho_c$ [\gcm] & $1.60 \pm 0.21$\\
Impact parameter b$_c$     & $0.910^{+0.036}_{-0.045}$ \\
Transit duration T$_{14,c}$ [h]   & $3.22^{+0.49}_{-0.30}$ \\
\hline
Planet g  &               \\
Orbital period P$_g$ [days]   &   $63.19 \pm 0.12$   \\
Time of inferior conj.$^b$ $T_{0,g}$ [BJD]$^a$      &     $8777.86^{+1.86}_{-1.82}$    \\
Orbital eccentricity $e_g$          & $0.055^{+0.024}_{-0.054}$  \\
Planet minimum mass M$_g$ [\Mearth]       & $5.75^{+0.76}_{-0.83}$  \\
\hline
Planet d  &              \\
Orbital period P$_d$ [days]    &   $278.3618 \pm 0.0003$ \\
Mid-transit time T$_{0,d}$ [BJD]$^a$     &    $7166.261 ^{+0.00061}_{-0.00057}$     \\
Orbital eccentricity $e_d$         & $0.063^{+0.027}_{-0.063}$  \\
Planet mass M$_d$ [\Mearth]      & $6.53^{+2.65}_{-3.15}$   \\
Planet radius $R_d$ [\Rearth]   &  $3.65 \pm 0.029$  \\
Bulk density $\rho_d$ [\gcm] & $0.74 \pm 0.33$\\
Impact parameter b$_d$        & $0.54 \pm 0.03$ \\
Transit duration T$_{14,d}$ [h]   & $12.71^{+0.46}_{-0.39}$ \\
\hline
Planet e  &             \\
Orbital period P$_e$ [days]   &  $393_{-5}^{+3}$  \\
Mid-transit time T$_{0,e}$ [BJD]$^a$    &    $7142.018^{+0.00051}_{-0.00048}$      \\
Orbital eccentricity $e_e$      & $0.065_{+0.031}^{-0.065}$  \\
Planet mass M$_e$ [\Mearth]      & $7.62^{+3.20}_{-4.63}$  \\
Planet radius $R_e$ [\Rearth]   & $5.19 \pm 0.04$ \\
Bulk density $\rho_d$ [\gcm] & $0.30^{+0.13}_{-0.18}$\\
Impact parameter b$_e$     & $0.632^{+0.034}_{-0.039}$ \\
Transit duration T$_{14,e}$ [h]  & $13.20^{+0.33}_{-0.26}$\\
\hline
Planet f  &              \\
Orbital period P$_f$ [days]    &  $542.0797 \pm 0.0001$  \\
Mid-transit time T$_{0,f}$ [BJD]$^a$      &   $7186.914^{+0.00018}_{-0.00022}$      \\
Orbital eccentricity $e_f$       & $0.052^{+0.025}_{-0.052}$  \\
Planet mass M$_f$ [\Mearth]      & $25 \pm 5$  \\
Planet radius $R_f$ [\Rearth]   &  $9.47 \pm 0.07$  \\
Bulk density $\rho_f$ [\gcm] & $0.16^{+0.03}_{-0.04}$\\
Impact parameter b$_f$     & $0.207^{+0.034}_{-0.031}$ \\
Transit duration T$_{14,f}$ [h]   & $18.9^{+0.85}_{-0.96}$ \\
\hline
Planet h (candidate)  &          \\
Orbital period P$_h$ [days] &  $2602_{-433}^{+468}$   \\
Time of inferior conj.$^b$ $T_{0,h}$ [BJD]$^a$     &    $8291^{+239}_{-225}$      \\
Orbital eccentricity $e_h$        & $0.066^{+0.033}_{-0.066}$   \\
Planet minimum mass M$_h$ [\Mearth]    & $43^{+16}_{-13}$ \\
\hline
\end{tabular}
\tablefoot{$^a$ BJD TBD - 2450000 -- $^b$ Time of inferior conjunction. Posteriors are indicated with median and $68\%$ highest density intervals. The full table with priors is available in the appendices.}
\label{tab_parameters}
\end{table}

\section{Discussion}
\label{sec:discussion}

\subsection{Architecture of the HIP~41378's system}
\label{sec:architecture}
The star HIP~41378 hosts a total of five transiting planets, a confirmed non-transiting one, and a candidate planet at very long orbital period. The system appears to be composed of two dynamically distinct sub-systems: an inner system with planets b, c, and g close to a 1:2:4 MMR and an outer system with planets d, e, and f. This separation is also reflected in the planetary inclinations. Planet b has an inclination of $i_b = 88.847 \pm 0.048$ degrees, and planet c has an inclination of $i_c = 88.475 \pm 0.013$ degrees. Assuming planet g is coplanar to planets b and c with an inclination close to 88.5\deg, we can compute its expected impact parameter knowing its semi-major axis of $0.3307 \pm 0.0023$. Using $b = a\cos{i}/R_*$, we find $b\approx1.05 > 1$, implying that planet g does not transit the stellar disc and is not compatible with any transits observed for this planet. The outer system, in contrast, appears to be tilted by roughly one degree relative to the inner system, with inclinations for planets d, e, and f between 89.8 and 89.9 degrees. We find no significant RV signal between the planet g (P$_g$=63 d) and planet d (P$_d$=278 d), resulting in an upper-mass limit of about 4-5\Mearth (see Fig. \ref{fig_detection_limit}). The TTVs of planets b and c were fully explained by the presence of planet g \citep{Leonardi2025}. Hence, the existence of two slightly non-coplanar sub-systems could account for the observed gap in planetary periods between 63 and 278 days.

\subsubsection{The orbital period of HIP 41378 d}

Our RV analysis favours an orbital period of $P_d = 278.3618 \pm 0.0003$ days for HIP~41378~d, consistent with one of the three transit-based solutions proposed to date \citep{Sulis2024}. The period is recovered independently in all RV data subsets and is statistically preferred by cross-validation model comparison. However, the non-detection of a transit by CHEOPS in 2023 \citep{Sulis2024} remains puzzling.
A possible explanation is the presence of large TTVs, as already observed for HIP~41378~f \citep{Bryant2021} and suggested for HIP~41378~d \citep{Sulis2024}. Dynamical interactions with planets e and f could shift transit times by several hours and up to several days, which is enough to move the expected transit outside the CHEOPS observation window. The TTV analysis remains limited by the small number of observed transits: only seven transits of planet f were detected \citep[Leonardi et al., in prep.]{Garcia-Mejia2025}. Only two photometric epochs have been secured for planet d, as well as a third tentative detection in spectroscopy \citep{Grouffal2022}. Without a new transit, $P_d$ cannot be definitively confirmed, although periods near 371 days are strongly disfavoured in the RV analysis.
Future transit monitoring with CHEOPS could detect additional transits---enabling a full TTV analysis and a definitive orbital solution allowing for follow-up observation---including transmission spectroscopy to compare with planet f \citep{Alam2022}.

\subsubsection{The orbital period of HIP 41378 e}

\begin{figure}[h]
                \centering
                \includegraphics[width=1\columnwidth]{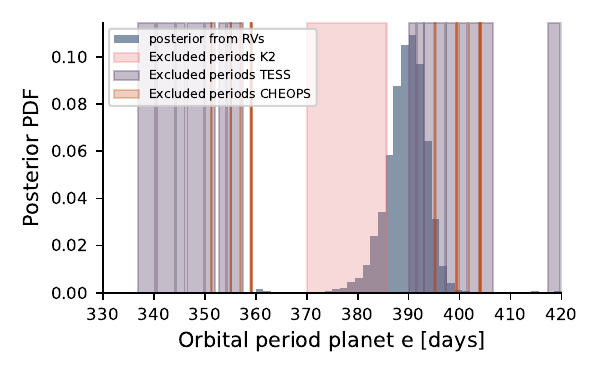}
                   \caption{Possible orbital periods for HIP~41378~e. The posterior distribution of the orbital period of planet e, obtained from the RV-only analysis with a broad uniform prior, is shown as a blue histogram. Period ranges excluded by K2 (pink), TESS (purple), and CHEOPS (orange) photometric observations are indicated as pink shaded areas.}
                          \label{fig_periods_e}
\end{figure}

Combining the RV data with K2 photometry yields a most probable orbital period of $P_e = 393^{+4}_{-3}$ days. As shown in Fig.~\ref{fig_periods_e}, most periods between 370 and 410 days are excluded by K2, TESS, and CHEOPS, leaving a narrow viable range at around 386–390 days, implying a missed transit shortly after K2 Campaign 18.

Alternative aliases appear in the nested sampling analysis (Appendix~\ref{appendice:nested_sampling}), notably near $\sim$190 and $\sim$760 days, representing half and twice the favoured period, respectively. These likely arise from harmonics or sampling effects of the $\sim$390-day signal. Periods of nearly 200 days were additionally ruled out by TESS. 

Another signal at $\sim$ 146 days was detected in the $\ell_1$ periodogram analysis (Section \ref{sec:l1}) and has been proposed to complete a near-three-body resonance configuration for all triplets of planets (Section \ref{sec:resonances}). While the RV phase coverage for a 146-day period is complete, attempting to place planet e at this period using photometric timing did not yield a converging solution. Using a uniform prior of between 100 and 250 days for planet e in the nested sampling analysis, we did not recover a signal near 146 days. The origin of this 146-day signal thus remains unclear and may correspond to another non-transiting planet. Model comparison continues to favour $P_e = 393^{+4}_{-3}$ days, although incomplete phase coverage and possible yearly systematics warrant further RV monitoring.

This estimate neglects potential TTVs. Given the large TTVs observed for planets d and f, similar variations for e could shift transit times by several hours, reopening some photometrically excluded windows. A definitive confirmation of the orbital period will therefore require either a dedicated transit search or a future joint RV–photometry analysis that includes TTVs, once additional transits of planets~d and~f become available. 

The favoured period lies close to a 5:7 commensurability, with both d and f and exact resonance implying $P_e \sim 388.45$ days. Within the 386d--390 day interval, the next two opportunities to observe a transit of planet e should occur between BJD=2461002 and BJD=2461042 (or near BJD=2461414.9 assuming exact resonance), and between BJD=2461388 and BJD=2461432 (near BJD=2461803.4 in the resonance case). 

Given that HIP~41378~e has a transit depth of 1.2~mmag and duration of 13.2 hours, any ground-based observatory that can secure nightly observations with sub-millimag precision, such as NGTS \citep{Bryant2021}, has a $>50$ \% probability of detecting one in-transit observation. Such a detection would significantly refine the ephemeris and enable a targeted space-based follow-up with, for example, CHEOPS. This would constitute only the second observed transit of planet~e since the event detected by K2 in 2015. Only a future transit will unambiguously confirm the precise orbital period of HIP~41378~e.

\subsubsection{Analysis of the resonances and possible additional planets}
\label{sec:resonances}
\begin{figure}[h]
\centering
\includegraphics[width=0.8\columnwidth]{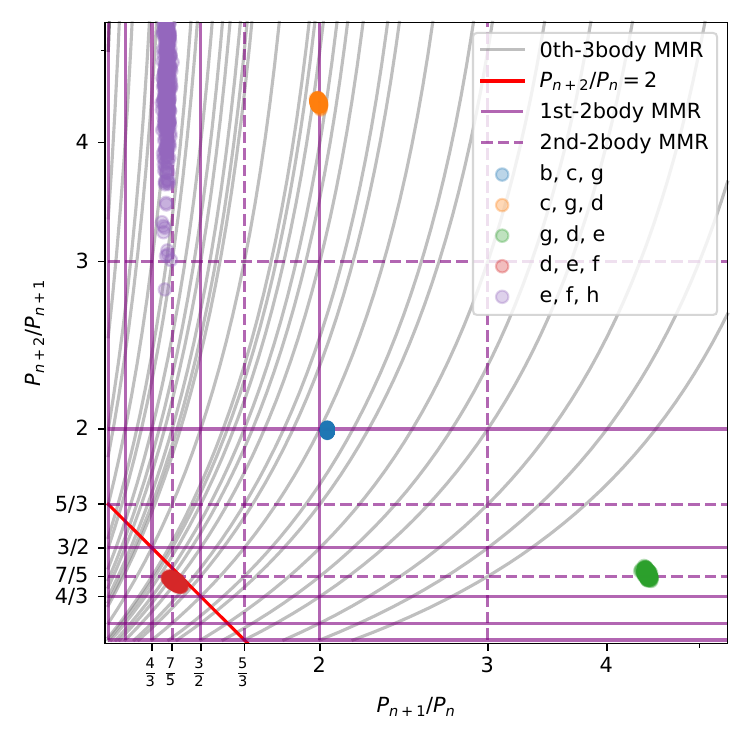}

\includegraphics[width=0.8\columnwidth]{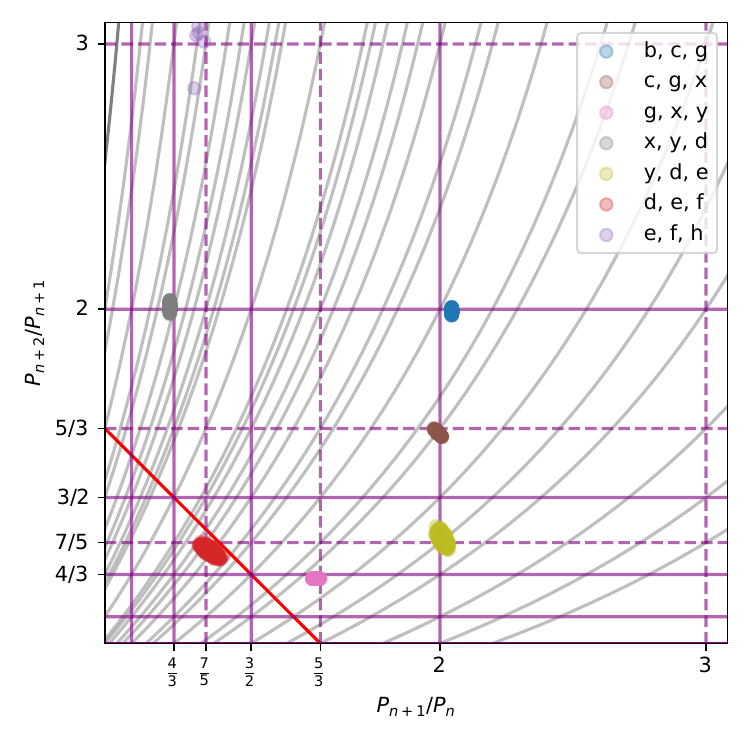}
   \caption{Subsequent triplet of HIP~41378 system with respect to MMRs. The solid purple lines show first-order two-planet MMRs, the dashed purple lines show second-order two-body MMRs, the solid red line shows the 2:1 MMR between planet $n$ and $n+2$, and the grey lines show zeroth-order three-body MMRs of the form $k/P_{n} -(k+q)/P{n+1} +q/P_{n+2}=0$ \citep[e.g.][]{Cerioni2022}. The top panel shows the RV posterior of the known planets. The bottom panel shows the example of case (iii) - solution 1, for which we added two planets, x and y, in order to reconnect planets c to f in a resonant chain.}
      \label{MMR}
\end{figure}

Using the RV posteriors, we examined the proximity of the system to two- and three-body MMR (Fig.~\ref{MMR}). The inner c–g pair is consistent with a 2:1 commensurability, while b–c lies near to but not inside resonance \citep{Leonardi2025}.

The compact d–e–f triplet is compatible with several three-body resonances, notably a chain of second-order 7:5 MMRs. In contrast, the gap in period ratios between the pairs g-d and f-h results in three sub-systems (c-b-g, d-e-f, and h) that do not have strong (i.e. resonant) gravitational interactions. We note, however, that the current posterior of h makes a 3:1 MMR with planet f possible.

Given the relatively loose constraints on h, we now focus on the rest of the system and see how missed planets could dynamically reconnect the inner and outer triplets of known planets considering that the period ratios between g and d is $\approx 4.4$. Following \cite{Leleu2021} and \cite{Luque2023}, we minimised the distance to two and three-body MMRs when adding planets. We discuss three cases, which are listed below: (i) one additional planet, minimising the distance to two-body MMRs; (ii) one additional planet, minimising the distance to three-body MMRs; and (iii) two additional planets, minimising distance to two-body MMRs.

(i) The gap between g and d could be filled by an additional planet, resulting in a 3:2 MMR followed by a 3:1 MMR, with a planet x at $\sim 95$ days. Equivalently, 3:1 followed by 3:2 works, with a planet x at $\sim 188$ days; this is the case shown in Figure \ref{MMR_p1}. 

(ii) A slightly more dynamically compact solution can be found by putting a planet x at $\sim 148$ days, resulting in a near-three-body resonance for all triplets involving said planet x; however, this results in pairs g--x and x--d at period ratios of 2.34 and 1.88, respectively, which are hence far from significant two-body MMRs. This is the case shown in Figure \ref{MMR_p1_3b}. 

(iii) Adding two planets allowed us to fully reconnect the system through a chain of two-body MMRs (5:3, 4:3, 2:1), leading to six equivalent solutions (Table~\ref{tab_MMR}); one is shown in Fig.~\ref{MMR}.

\begin{table}
\caption{Possible combination of periods for hypothetical planets x and y to reconnect HIP41378 as a resonant chain.}
\begin{tabular}{c|ccccccc}

 solution & 1  & 2 & 3 & 4 & 5 &6 \\ 
 \hline
$P_x$ [d] & 126.4 & 126.4 & 105.3 & 105.3 & 84.3  & 84.3 \\
$P_y$ [d] & 210.7 & 168.5 & 210.7 & 140.4 & 168.5 & 140.4 \\
\end{tabular}
\label{tab_MMR}
\end{table}

The near-resonant nature of b–c–g and d–e–f motivated this search. Consecutive second-order 7:5 MMRs, as suggested for d–e–f, are rare (only TOI-1136 shows a similar case; \citealt{Dai2023}). While cases (i) and (ii) do not strictly lie inside two-body resonances, this is not necessarily problematic, as b–c itself is slightly offset, and HIP 41378 may represent a partially disrupted chain. Case (iii) produces compact resonant configurations comparable to other known chains, consistent with the system’s puffy and nearly coplanar architecture \citep{Leleu2024b}.

\subsubsection{A signal at very long orbital period}

We detected a low-amplitude, long-period signal (P$_h$ = $2602_{-433}^{+468}$) in the RV data, which is also present, albeit less clearly and at different orbital periods, in some stellar activity indicators. This signal is primarily driven by the ESPRESSO dataset (see Figure~\ref{fig_posterior_priors_RV}), particularly the third observing campaign (November 2022--May 2023). The phase of this signal does not match any known transiting planet in the system, and it rules out planets d and e.

We considered the possibility of an instrumental origin. However, no such offset is seen in ESPRESSO constant stars \citep{Figueira2025}, nor in the activity indicators at the relevant level. This suggests a physical origin for the signal.

An alternative explanation is that this signal may be related to a stellar magnetic cycle. Some activity indicators (e.g. FWHM, CCF contrast, S-index) show low-amplitude variations on timescales of 900–1700 days, which is roughly compatible with the period of the RV signal. However, modelling the RVs jointly with these indicators using a squared-exponential GP did not converge, and the signal in the RVs remains significant, independent of activity trends.

Its origin therefore remains uncertain: it may correspond to a long-period planet candidate, a stellar magnetic cycle, or, most plausibly, a combination of both. We tentatively modelled it as a Keplerian signal but stress that its planetary nature is not established. Continued long-term RV monitoring and activity characterisation are required. Future Gaia DR4/DR5 time-series astrometry may detect an acceleration or orbital motion and, combined with the RV dataset, help constrain the mass and nature of this companion.

\subsubsection{Total mass of the planetary system}

The HIP~41378 system hosts six planets up to $\sim 1.5$ AU. Given the observed resonances in the inner and outer sub-systems (see Sections \ref{sec:architecture} and \ref{sec:resonances}), the architecture is likely complete up to the orbit of planet f. The most likely missed planet would be between the planets d and g, but would be less massive than a few Earth masses (see Fig. \ref{fig_detection_limit}). The total planetary mass of the system up to 1.5 AU is of $\sim$ 58 \Mearth. This value is substantially greater than that of the Solar System ($\sim$ 2 \Mearth), which might be explained by a more massive disc of a more massive host star. However, if we integrate the total planetary mass of the Solar System, up to the orbit of Neptune ($\sim$ 450 \Mearth), HIP~41378 is missing several hundreds of \Mearth, even including planet h. We thus speculate that HIP~41378 hosts some massive planets beyond 1.5 AU. Given the detection limits (see Fig. \ref{fig_detection_limit}), Jupiter-mass planets could be compatible with our decade-long RV time series if their orbital periods exceed $\sim$ 20000 days (55 years). Detecting these potential outermost and massive planets would require substantially longer RV time baselines. These companions might also be probed by next-generation high-contrast imaging instruments such as the Habitable Worlds Observatory \citep[HWO,][]{HWO2021} and the Planetary Camera and Spectrograph \citep[PCS,][]{Kasper2021} on the Extremely Large Telescope (ELT), although likely at the limit of detectability. A Jupiter-like planet around HIP~41378 with a 55-year orbital period would have an angular separation of about 0.13'' and would be easily resolved. The expected reflected-light contrast, of the order of a few $\times 10^{-10}$, could be accessible with HWO considering the predicted performances \citep{Feinberg2026}. Alternatively, HIP~41378 and the Solar System might have formation mechanisms of different efficiency to form massive planets, especially as the former is a slightly metal-poor star. 

\subsection{The nature of HIP~41378~f}

\begin{figure}[h!]
                \centering
                \includegraphics[width=1.0\columnwidth]{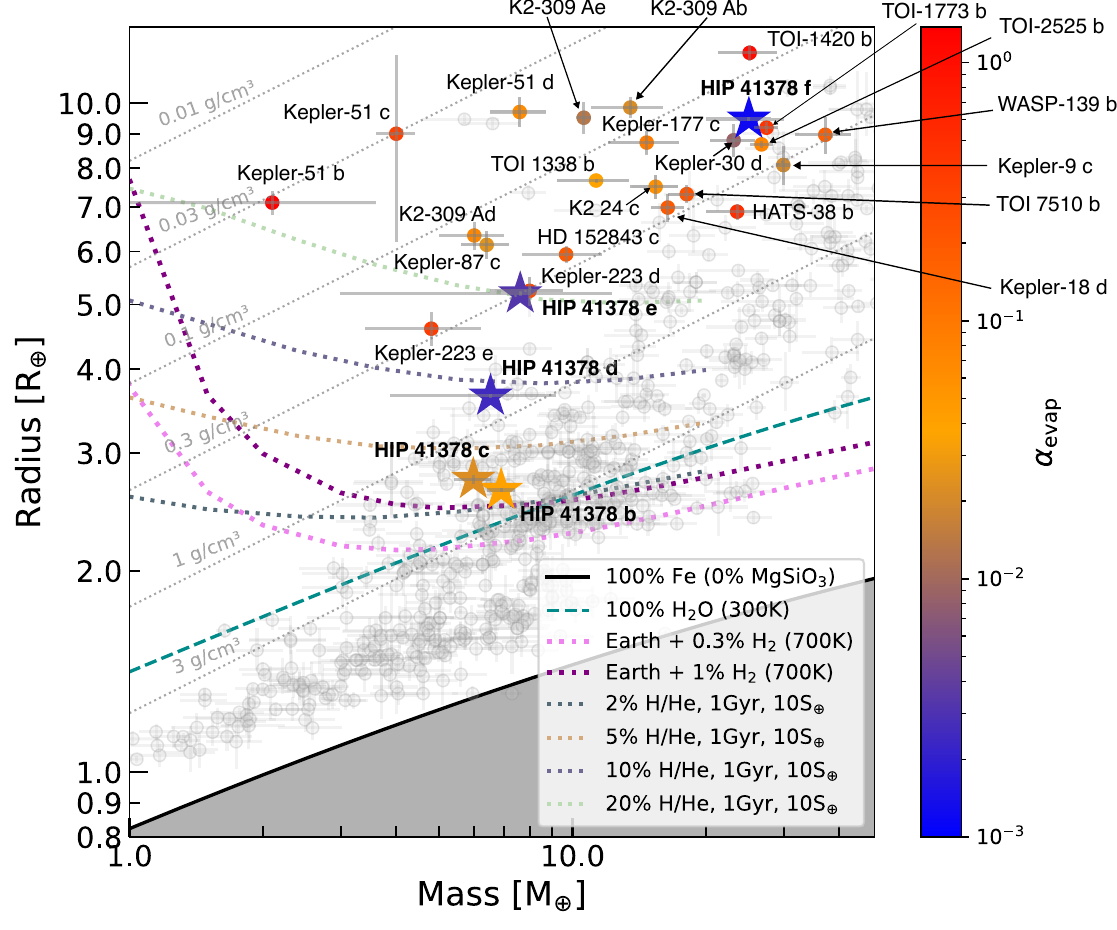}
                   \caption{Mass-radius diagram for HIP~41378 transiting planets. Confirmed planets from the \href{https://exoplanetarchive.ipac.caltech.edu/}{NASA Exoplanet Archive} with 20$\%$ precision on mass and radius are represented as grey dots. Planets with densities under 0.2\,g/cm$^{-3}$ are highlighted. The colours show the importance of mass loss through $\alpha_{\rm evap}$. Theoretical composition models for rocky planets and water worlds from \citet{Zeng2019} are represented as dashed lines, and rocky planets with H/He atmospheres from \citet{Lopez2014} are represented as dotted lines. The isodensity curves are show straight lines. Parts of this plot were prepared using \href{https://github.com/castro-gzlz/mr-plotter/}{mr-plotter} \citep{Castro-Gonzales2023}.}
                          \label{fig_mass_radius}
\end{figure}

The five transiting planets of HIP~41378 span a wide range of orbital distances, yet they show a trend of decreasing density with radius and with distance from the star. The mass--radius diagram is represented in Figure \ref{fig_mass_radius}. Planets b and c are consistent with sub-Neptunes with moderate H/He envelopes, while planets d and e are less dense. 

For HIP~41378~f, our revised mass of $25 \pm 5 M_\oplus$ is twice as large as the $12 \pm 3 M_\oplus$ reported in \citet{Santerne2019}. Its measured density is now estimated as $\rho_f = 0.16^{+0.03}_{-0.04}$ g\,cm$^{-3}$. The planet still belongs to the class of super-puff planets. 

\subsubsection{The planet's low density}

The low density of HIP~41378~f ($\rho_f = 0.16^{+0.03}_{-0.04}$ g\,cm$^{-3}$) is among the lowest measured for exoplanets and may at first appear difficult to reconcile with standard planet formation models. Indeed, conventional theories indicate that dense Earth-like cores in protoplanetary discs start accreting efficiently when the core mass is in excess of about $\rm 10 M_\oplus$ \citep{Mizuno1980, Pollack1996}, leading to planetary densities much in excess of the measured value. However, when both accretion rates and opacities are reduced compared to traditional values, much smaller cores, even down to about $\sim 1-5 M_\oplus$, can become supercritical and accrete a significant hydrogen--helium envelope within the lifetime of the discs \citep{Ikoma2000}. 

The first example of planets with extremely low densities, also known as super-puffs, are the planets in the Kepler-51 system that all range between 0.03 and $0.05\rm g/cm^3$ \citep{Masuda2014, Masuda2024}. As shown by \citet{Lee2016} these planets may form at large orbital distances in quiescent discs and migrate, potentially retaining their large radius and small densities. Their low density implies that the most stringent limitation for their survival may be resisting to the XUV stellar flux that they receive and the subsequent mass loss \citep{Lopez2014}. 

Figure~\ref{fig_mass_radius} shows a mass-radius diagram focusing on low-density planets with precise masses and radii. The colours correspond to the values of a mass-loss parameter $\alpha_{\rm evap}$ defined as the ratio of the time integral of the UV flux received by the planet to its binding energy: 
\begin{equation}
\alpha_{\rm evap} \equiv E_{\rm UV} / E_{\rm bind},
\end{equation}
where 
\begin{equation}
E_{\rm UV} = \frac{\pi R_{\rm p}^2}{a^2}\int_0^t F_{\rm UV} dt,
\end{equation}
$R_{\rm p}$ being the planetary radius, $a$ its orbital semi-major axis, and $F_{\rm UV}(t)$  the UV flux of the star, which was taken as equal to 29.7 (age/Gyr)$^{-1.23}$\,erg/s/cm$^2$ following \citet{Ribas2005}. We also accounted for a saturation of that relation for ages of < 100 Myr. The relation is formally valid only for solar-type stars, and we neglected the mass dependence (stars highlighted in Fig. \ref{fig_mass_radius} range from 0.69 to 1.25\,M$_\odot$).  The binding energy is approximated as $E_{\rm bind} = \frac{G M_{\rm p}^2}{R_{\rm p}}$, where $G$ is the gravitational constant and $M_{\rm p}$ the planetary mass. 

As discussed first in a slightly different form by \citet{Lecavelier2007} and by \citet{Lopez2014}, $\alpha_{\rm evap}$ indicates the magnitude of mass loss due to XUV radiation. We expect $\alpha_{\rm evap} < 1$ and this is mostly what we find for the planets shown in Fig.~\ref{fig_mass_radius}. For the Kepler-51 planets, we obtain $\alpha_{\rm evap}$ values of 1.5, 0.4, and 0.08 for planets b, c, and d. Planet b appears most vulnerable to atmospheric escape, although uncertainties in its mass may bring $\alpha_{\rm evap}$ below unity. Given the young age of the system ($0.7\pm0.5$ Gyr; \citealt{Masuda2024}), continued mass loss may still alter their bulk properties.

In contrast, the planets orbiting HIP~41378 exhibit significantly smaller values of $\alpha_{\rm evap}$. The largest value in the system is 0.04 (planet b), while HIP 41378 f has $\alpha_{\rm evap}=0.001$. The large orbital separation of planet f and the system’s age of $\sim2$ Gyr imply that atmospheric escape has likely played only a minor role in shaping its present-day structure.

Among the low-density planets occupying a similar region of the mass–radius diagram as HIP 41378 f are planets b and e around the young star K2-309 A (V1298 Tau), a 20 to 30 Myr, 1.1\,M$_\odot$ star. They are characterised by $\alpha_{\rm evap}$ = 0.02 and 0.013, implying that they should suffer more mass loss over the lifetime of the star. \citet{Karalis2025} showed that the planets’ densities may be accounted for with a gas-to-core ratio that approaches unity. They probably migrated from beyond 1\,au before reaching their present-day orbits at 0.17 and 0.27\,au, respectively. 

Overall, while the low density of HIP 41378 f is unusual, our analysis indicates that its envelope is energetically stable against XUV-driven escape over the system’s lifetime. Its current structure, therefore, does not require extreme ongoing mass loss to be explained.

As we detail in Section \ref{sec:rings}, the presence of planetary rings remains a possible interpretation of the large measured radius of HIP~41378~f. However, alternative formation pathways must be studied in detail. The possibility that incoming planetesimals and pebbles may be filtered by other planetary companions at larger orbital distances \citep[e.g.][]{Guillot2014} may help us to understand their formation. Detailed studies of these kinds of planets and a characterisation of their atmospheres would allow a much better understanding of the growth of planetary cores and the importance of neighbouring planetary companions during the process. 

\subsubsection{The ring hypothesis}
\label{sec:rings}

A full understanding of the formation pathway of HIP 41378 f is essential to explain its low density. We therefore revisited the alternative ring scenario proposed by \citet{Akinsanmi2020} and assessed its compatibility with the updated system parameters. This hypothesis remains viable, although it requires a somewhat higher intrinsic density than initially assumed. Adopting $R_\text{true}=3.9$~$R_\oplus$ as the true planet's radius yields a true density for planet~$f$ of $2.3$~$\text{g\,cm}^{-3}$ and an observed radius of $R_f\approx 2.4$~$R_\text{true}$, which is close to the Roche limit, as expected for a Saturn-like tidal ring. Using Eq.~1 of \citet{Akinsanmi2020}, we find the required ring particle density is $2.47$~g.cm$^{-3}$, which is consistent with a rocky composition and long-term survival at the planet’s equilibrium temperature (294 K).

\begin{figure}
   \includegraphics[width=\columnwidth]{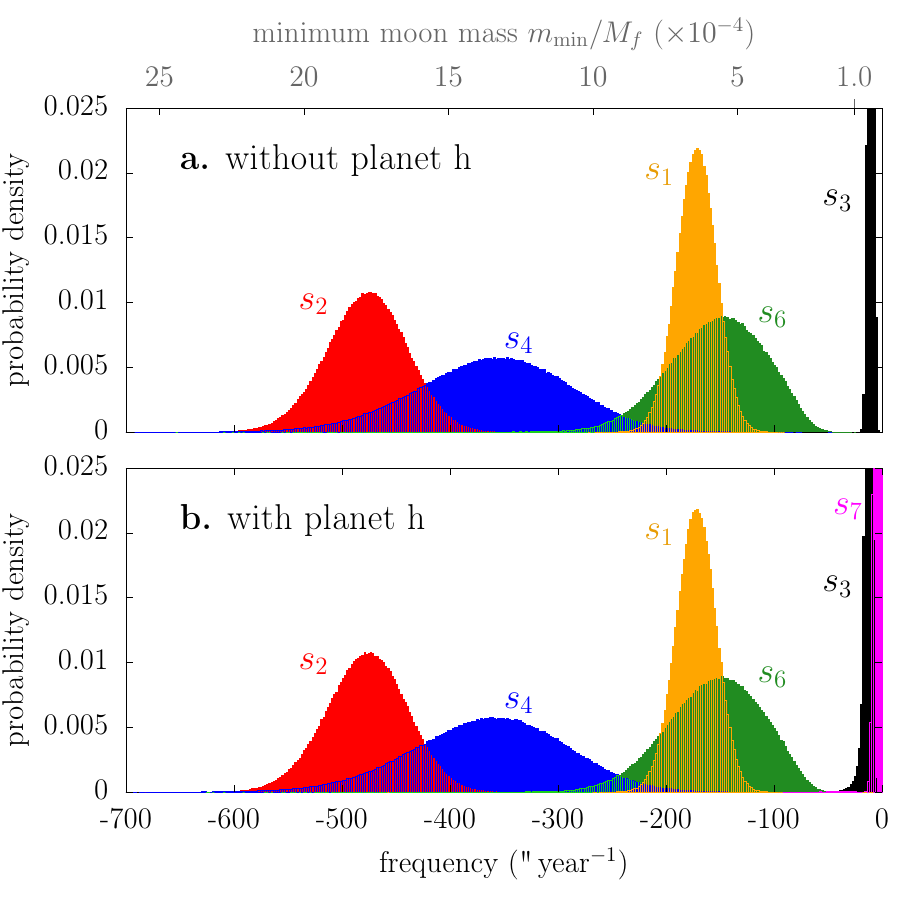}
   \caption{Probability density of the inclination proper modes of the HIP\,41378 system with and without the candidate planet~h. Histograms are built from $10^6$ realisations of the Lagrange-Laplace system with the mass and semi-major axis uncertainties listed in Table~\ref{tab_parameters}. For planets~g and~h, the minimum mass is taken as the true mass. The histograms for frequencies $s_3$ and $s_7$ peak above the top border of the graph. Frequency $s_5$ is identically equal to zero from the conservation of angular momentum. The upper axis shows the minimum moon mass needed for planet~$f$ to be fully tilted through a resonance with a given frequency value $s$.}
   \label{fig_LLfreq}
\end{figure}

As shown by \cite{Akinsanmi2020}, if a ring is present, it must lie close to the sky plane, implying an obliquity of about $90^\circ$, which is similar to that of Uranus. Such a tilt requires a dynamical mechanism, as proposed by \citet{Saillenfest2023} and \citet{Lu2025}, but it remains to be verified that these mechanisms are still compatible with the updated system parameters in Table~\ref{tab_parameters}. These mechanisms rely on the existence of resonances between the nodal precession modes of the planet's orbital plane (produced by the planet-planet gravitational interactions) and its spin-axis precession (produced by the stellar torque applied on the planet's equatorial bulge). The precession modes of planet~$f$'s orbital plane depend on the masses and semi-major axes of all planets in the system. A good estimate for the dominant modes is given by the Lagrange--Laplace theory. Figure~\ref{fig_LLfreq} shows the probability density of these modes considering the parameters from Table~\ref{tab_parameters} and their uncertainties. By comparing Fig.~\ref{fig_LLfreq} with Fig.~3 of \cite{Saillenfest2023}, we note that the updated parameters decrease the uncertainties of all frequencies, especially frequency $s_4$, without producing any major rearrangement in the system.

The inclusion of the distant candidate planet~h only adds a very low-frequency term ($s_7$), corresponding to a slow precession of the inner system around the total angular momentum.

In the scenario of \cite{Saillenfest2023}, a migrating moon around planet~$f$ gradually changes its spin-axis precession frequency until it matches one of the modes ($s)$ shown in Fig.~\ref{fig_LLfreq}. At that point, a resonance is triggered, and the planet's spin axis is tilted on a gigayear timescale up to an obliquity of about $90^\circ$, and it eventually leads to tidal disruption of the moon and ring formation. This mechanism has been shown to be generic (see e.g. \citealp{Saillenfest2020,Saillenfest2021,Saillenfest2022}). The top horizontal axis in Fig.~\ref{fig_LLfreq} shows the minimum mass, $m_\mathrm{min}$, that the moon must have had in order to tilt planet~$f$ through a resonance with a frequency of $s$ along the bottom horizontal axis. The values of $m_\mathrm{min}$ were computed using Eq.~(5) of \cite{Saillenfest2023}, the assumed radius $R_\text{true} = 3.9$~$R_\oplus$, and the spin rate and oblateness of Uranus. If $R_\text{true}$ is taken as $R_f$ instead, then the values obtained for $m_\mathrm{min}$ are multiplied by roughly a factor of three. Figure~\ref{fig_LLfreq} shows that in order to be tilted through a resonance with mode $s_6$, planet~$f$ must have had a moon with mass ratio of $m/M_f$ equal to a few $10^{-4}$. This is consistent with what is expected for regular moons around giant planets (see \citealp{Canup2006}, \citealp{Saillenfest2023}). For a 25 \Mearth\ planet, this would correspond to a moon of $\sim 0.005 - 0.01$ \Mearth, which is below the maximum satellite mass of $\sim 0.15$ \Mearth\ found to be stable over long timescales for HIP~41378 by \citet{Harada2023}. Although such a moon would be undetectable, it would be sufficient to drive the mechanism.

The mechanism of \cite{Lu2025} is similar, except that it relies on the early disc-driven planetary migration in order to trigger the resonance, and it does not explain the existence of the ring. If we consider that planet~$f$ did not have a moon during the planetary migration phase, its spin-axis precession frequency would have been very slow (see Eqs.~6 and 7 of \citealp{Lu2025}); yet, Fig.~\ref{fig_LLfreq} shows that frequencies $s_3$ and $s_7$ extend to very low absolute values, still allowing for resonances with planet~$f$'s slow spin-axis precession. If instead planet~$f$ had a moon during planetary migration, Fig.~\ref{fig_LLfreq} shows that a resonance with frequency $s_6$ would also be reachable. Therefore, the mechanism proposed by \cite{Lu2025} also remains a viable mechanism when considering the updated parameters for the HIP\,41378 system given in Table~\ref{tab_parameters}.

\section{Conclusion}
\label{sec:conclusion}

Following the first set of transits detected by K2 in 2015, we monitored the HIP~41378 system from 2016 to 2025 in RV. Over 1000 spectra were obtained in one decade over 621 individual nights, mainly with the HARPS, HARPS-N, HIRES, and ESPRESSO spectrographs. The analysis of these data, combined with the K2 photometry and constraints from TESS and CHEOPS, allowed us to determine the orbital periods and masses of six planets in the system (ordered by increasing distance from the host star): b, c, g, d, e, f. The RV amplitudes range from 0.5 \ms\ to 1.8 \ms\, with typical precision levels between 6\% and 50\%. 

Compared to previous studies on this system, the present work provides several improvements. These are listed below.
\begin{itemize}
    \item We find strong evidence that planets d and e have orbital periods of $P_d \sim 278$ days and $P_e \sim 386$–390 days.
    \item We refined the masses for all planets, including planets d and e, and doubled the mass estimate for the super-puff planet~f.
    \item We detected a long-period planet candidate, HIP~41378~h, of $m \sin i \sim 43 M_\oplus$ and a period of six to eight years.
    \item We observe a clear trend of decreasing density with semi-major axis for the transiting planets of the system.
    \item We investigated the system architecture, which appears to consist of two dynamically distinct sub-systems. Additional undetected planets could exist and potentially connect the two sub-systems in a resonant chain.
\end{itemize}

However, there are still open questions concerning this system that are not fully addressed here and require further observations.

\begin{itemize}
    \item A third (second) photometric transit detection of planet d (e, respectively) is needed to fully secure their orbital period and derive an accurate ephemeris. The CHEOPS space telescope is perfectly suited for that, although the observation time span should last long enough to account for the large TTVs \citep{Sulis2024}.
    \item The nature of the low-density planet f is still uncertain. Its properties are consistent with either an inflated temperate planet or the presence of rings. Observing the transit in the near-infrared with, for example, the JWST or ARIEL, should allow us to characterise the planet's atmosphere and radius more accurately, thereby testing the 'inflating planet' scenario and confirm or rule out the ring hypothesis \citep{Alam2022}.
    \item The candidate planet h needs to be secured via long-term spectroscopic observations or a detailed characterisation of the stellar variability across one decade.
    \item The architecture of the system appears almost complete up to 1.5 AU and down to a few Earth masses, although one or two planets might still be undetected between planets g and d. Further observations could deeply explore this rich system. Comparison with the Solar System indicates that more massive planets might still be discovered beyond $\sim$ 50 AU, which would require next-generation high-contrast imaging instruments to be detected.
\end{itemize}

The diversity of planets' properties in this bright system, from warm sub-Neptunes to temperate inflated planets, makes HIP~41378 a unique laboratory for comparative planetology within the system and with the Solar System. However, this requires maintaining a precise ephemeris in the future to schedule follow-up observations. More transit observations of the outermost planets (Leonardi et al., in prep.) are thus needed in this regard.

\section{Data availability}
Radial velocity data are only available in electronic form at the CDS via anonymous ftp to cdsarc.u-strasbg.fr (130.79.128.5) or via http://cdsweb.u-strasbg.fr/cgi-bin/qcat?J/A+A/.

\begin{acknowledgements}
We acknowledge the contribution from the PFS team with their early observations of the system. We made use of the SIMBAD database, CDS, Strasbourg Astronomical Observatory, France and of the VizieR catalogue access tool, CDS, Strasbourg Astronomical Observatory, France (DOI : 10.26093/cds/vizier). The project leading to this publication has received funding from the Excellence Initiative of Aix-Marseille University - A*Midex, a French “Investissements d’Avenir programme” AMX-19-IET-013. This work was supported by the "Programme National de Planétologie" (PNP) of CNRS/INSU co-funded by CNES. This work was supported by the "Action Thématique Physique Stellaire” of CNRS/INSU PN ASTRO co-funded by CEA and CNES. This work is based on observations made with the Italian Telescopio Nazionale Galileo (TNG) operated on the island of La Palma by the Fundaci\'on Galileo Galilei of the INAF (Istituto Nazionale di Astrofisica) at the Spanish Observatorio del Roque de los Muchachos of the Instituto de Astrofisica de Canarias. The HARPS-N project was funded by the Prodex Program of the Swiss Space Office (SSO), the Harvard University Origin of Life Initiative (HUOLI), the Scottish Universities Physics Alliance (SUPA), the University of Geneva, the Smithsonian Astrophysical Observatory (SAO), the Italian National Astrophysical Institute (INAF), University of St. Andrews, Queen’s University Belfast, and University of Edinburgh. This work has made use of data from the European Space Agency (ESA) mission {\it Gaia} (\url{https://www.cosmos.esa.int/gaia}), processed by the {\it Gaia} Data Processing and Analysis Consortium (DPAC, \url{https://www.cosmos.esa.int/web/gaia/dpac/consortium}). Funding for the DPAC has been provided by national institutions, in particular the institutions participating in the {\it Gaia} Multilateral Agreement. The work of HPO has been carried out within the framework of the NCCR PlanetS supported by the Swiss National Science Foundation under grants 51NF40 182901 and 51NF40 205606. NCS is co-funded by the European Union (ERC, FIERCE, 101052347). Views and opinions expressed are however those of the author(s) only and do not necessarily reflect those of the European Union or the European Research Council. Neither the European Union nor the granting authority can be held responsible for them. This work was supported by FCT - Fundação para a Ciência e a Tecnologia through national funds by these grants: UIDB/04434/2020 DOI: 10.54499/UIDB/04434/2020, UIDP/04434/2020 DOI: 10.54499/UIDP/04434/2020. We acknowledge financial contribution from the INAF Large Grant 2023 ''EXODEMO''. This research was funded in part by the UKRI (Grants ST/X001121/1, EP/X027562/1). Some of the data presented herein were obtained at Keck Observatory, which is a private 501(c)3 non-profit organisation operated as a scientific partnership among the California Institute of Technology, the University of California, and the National Aeronautics and Space Administration. The Observatory was made possible by the generous financial support of the W. M. Keck Foundation. We recognise and acknowledge the cultural role and reverence that the summit of Maunakea has within the indigenous Hawaiian community. We are deeply grateful to have the opportunity to conduct observations from this mountain. A.M. acknowledges funding from a UKRI Future Leader Fellowship, grant number MR/X033244/1 and a UK Science and Technology Facilities Council (STFC) small grant ST/Y002334/1. S.C.C.B.acknowledges the support from Fundação para a Ciência e Tecnologia (FCT) in the form of work contract through the Scientific Employment Incentive program with reference 2023.06687.CEECIND and DOI 10.54499/2023.06687.CEECIND/CP2839/CT0002. P.F. acknowledges financial support from the Severo Ochoa grant CEX2021-001131-S funded by MCIN/AEI/10.13039/501100011033. P.F. is also funded by the European Union (ERC, THIRSTEE, 101164189). BA acknowledges the support of the Swiss National Science Foundation under grant number PCEFP2 194576. C.A.W. would like to acknowledge support from the UK Science and Technology Facilities Council (STFC, grant number ST/X00094X/1). X.D acknowledges the support from the European Research Council (ERC) under the European Union’s Horizon 2020 research and innovation programme (grant agreement SCORE No 851555) and from the Swiss National Science Foundation under the grant SPECTRE (No 200021 215200). AL acknowledges support of the Swiss National Science Foundation under grant number TMSGI2\_211697. ACC acknowledges support from STFC consolidated grant number ST/V000861/1, UKRI/ERC Synergy Grant EP/Z000181/1 (REVEAL), and UKSA grant number ST/X002217/1. O.D.S.D. acknowledges support from e-CHEOPS (PEA No 4000142255
\end{acknowledgements}

\bibliographystyle{aa} 
\bibliography{biblio}

\begin{appendix}

\section{Stellar activity}
\label{sec:stellar_activity}

Figure~\ref{fig_periodograms_activity} presents the time series and GLS periodograms of several stellar activity indicators: the CCF FWHM, BIS, CCF contrast, and the Ca~II H$\&$K S-index. These diagnostics are used to search for signatures of stellar rotation and long-term magnetic variability, and to assess their potential impact on the RV measurements.

Prior to computing the periodograms, we removed linear trends and inter-instrument offsets from each indicator. In the periodograms, we indicate the orbital periods of the transiting planets, as well as the stellar rotation periods inferred from $\log(R'_\mathrm{HK})$ and photometry. The highest peaks and their one-year aliases are also marked.
We additionally computed Spearman rank correlation coefficients between each activity indicator and the RV time series, both globally and per instrument, and found little to no correlation in all cases.

Below, we analyse each indicator in detail.

\subsection{FWHM}

The FWHM periodogram displays a dominant peak at $\sim1735$~days, with a secondary peak at $\sim463$~days corresponding to a one-year alias. The long-period signal can be well modelled by a sinusoid with a period of 1714 days. After subtracting this signal, no significant power remains at shorter periods that could be associated with stellar rotation.

\begin{figure}[h]
     \centering
     \begin{subfigure}[b]{0.45\textwidth}
         \includegraphics[width=\textwidth]{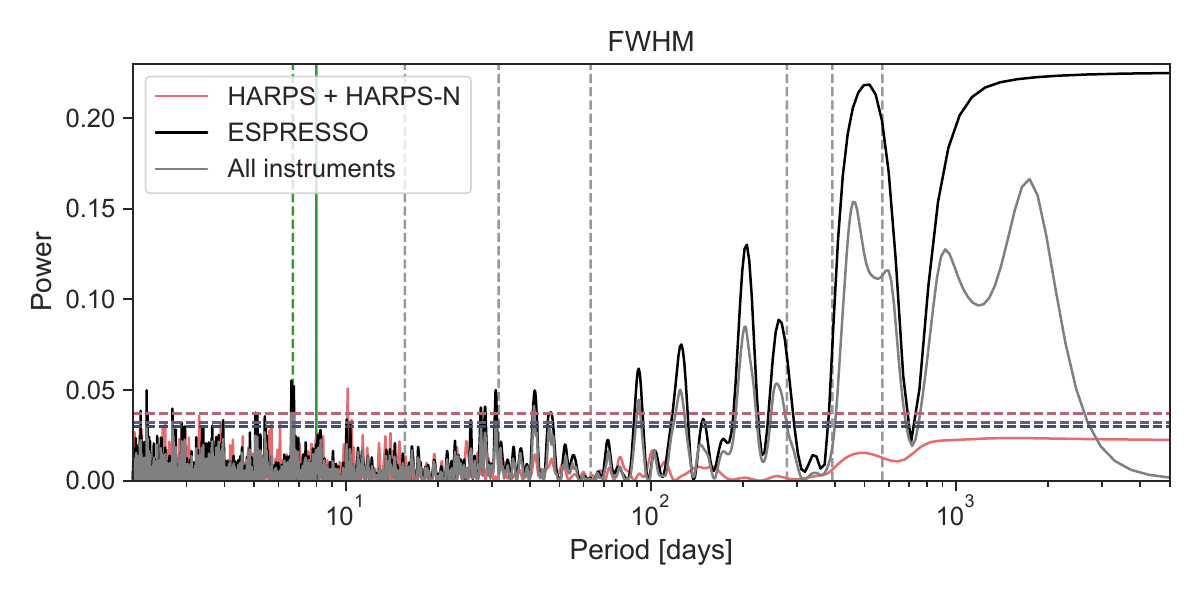}       
     \end{subfigure}
     \begin{subfigure}[b]{0.45\textwidth}
 
         \includegraphics[width=\textwidth]{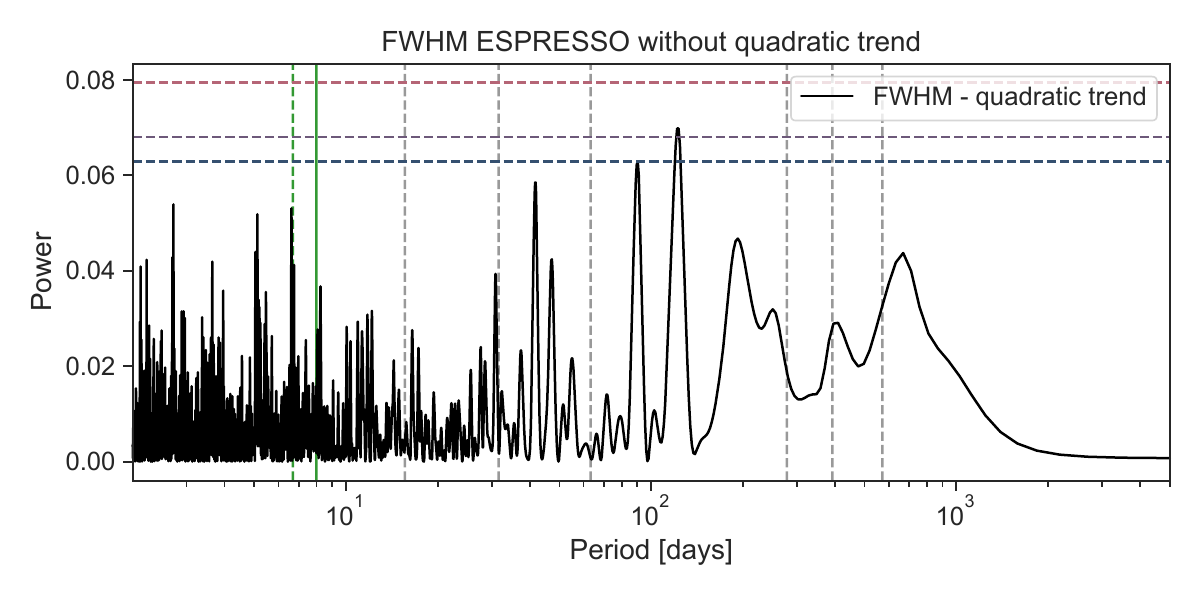}   
     \end{subfigure}
\caption{Frequency analysis of the FWHM. \textit{Upper panel:} Periodograms of the FWHM timeserie with only HARPS and HARPS-N data in pink, only ESPRESSO data in black and all instruments together in grey. The two rotation periods derived from $\log(R'_\mathrm{HK})$ and photometry are represented as, respectively, solid and dotted green lines. Vertical dotted black lines represent the period of the known planets in the system. \textit{Lower panel:} Periodogram of ESPRESSO data after removing a quadratic-trend. The horizontal dashed lines in the two panels represents the FAP levels at 0.1, 1 and 10$\%$.}
\label{fig_activity}
\end{figure}

This long-term variability is primarily driven by a quadratic trend visible in the ESPRESSO data. To illustrate this, we computed GLS periodograms separately for ESPRESSO and for HARPS/HARPS-N, as shown in Fig.~\ref{fig_activity}. Once the quadratic trend is removed from the ESPRESSO FWHM time series, no significant periodic signal is detected.

Therefore, the FWHM does not provide clear evidence for rotationally modulated stellar activity in this system. Moreover, the long-period signal detected in the FWHM (1714 days) is significantly different from the long-term signal identified in the RV time series, suggesting that it is unlikely to be responsible for the RV variability, or at least not in its entirety.

\subsection{BIS}

After removing a linear trend, the BIS periodogram does not exhibit any statistically significant periodic signal. It is the only activity indicator that does not display long-period signals. Low-amplitude peaks are present near 8~days, close to the expected stellar rotation period, and at one-day aliases, but their power remains well below significance thresholds.

To further investigate the possible presence of rotational modulation, we modelled the BIS time series using a GP with a quasi-periodic kernel (Eq.~\ref{eq:QP_kernel}), implemented with \texttt{tinyGP}. We adopted uniform priors on the rotation period (2–20 days), amplitude (0–5 \ms), exponential decay timescale (1–500 days), and $\Gamma$ (0.1–10 days), and included instrument-dependent offsets and jitter terms.

Despite exploring several prior configurations on the rotation period, the GP fit failed to converge to a stable solution. This lack of convergence is consistent with the absence of a coherent and sufficiently strong quasi-periodic signal in the BIS time series, and suggests that rotationally induced activity is not significantly detected in this indicator.

\subsection{Contrast}

The CCF contrast shows the strongest correlation with the RVs among the considered indicators, particularly in the ESPRESSO data. Its GLS periodogram exhibits a dominant peak at $\sim1013$~days, which may be indicative of a low-amplitude stellar magnetic cycle. This long-period signal is well reproduced by a sinusoidal model at the corresponding period.

After subtracting this long-term variability, several low-power peaks remain at shorter periods, but none reach statistical significance or coincide robustly with the expected rotation period. We attempted to model the contrast time series with a quasi-periodic GP, both before and after removal of the long-period sinusoidal component, using the same priors as for the BIS.

In all cases, the GP modelling suffered from poor convergence, preventing a reliable determination of the stellar rotation period. This again suggests that rotational modulation in the contrast is weak or incoherent over the observational baseline.

\subsection{S-index}
The Ca~II H$\&$K S-index periodogram shows a significant peak near 948~days, close to the long-period signal detected in the CCF contrast, and likely associated with the same magnetic activity cycle. A sinusoidal fit yields a period of $969 \pm 27$~days for this long-term variability.

Once this signal is removed, no significant power remains at shorter periods that could be attributed to stellar rotation. As for the other indicators, quasi-periodic GP modelling did not converge, further indicating the absence of a detectable rotationally modulated signal in the S-index.

Overall, none of the activity indicators show a robust periodic signal at the expected stellar rotation period, nor do they exhibit strong correlations with the RV time series. The long-term variability detected in the FWHM, contrast, and S-index may be attributed to a low-amplitude magnetic cycle, however, its characteristic period does not fully coincide with the long-period signal observed in the RVs, which is significantly longer.

\section{Residuals analysis}

 \begin{figure}[h]
		\centering
		\includegraphics[width=1\columnwidth]{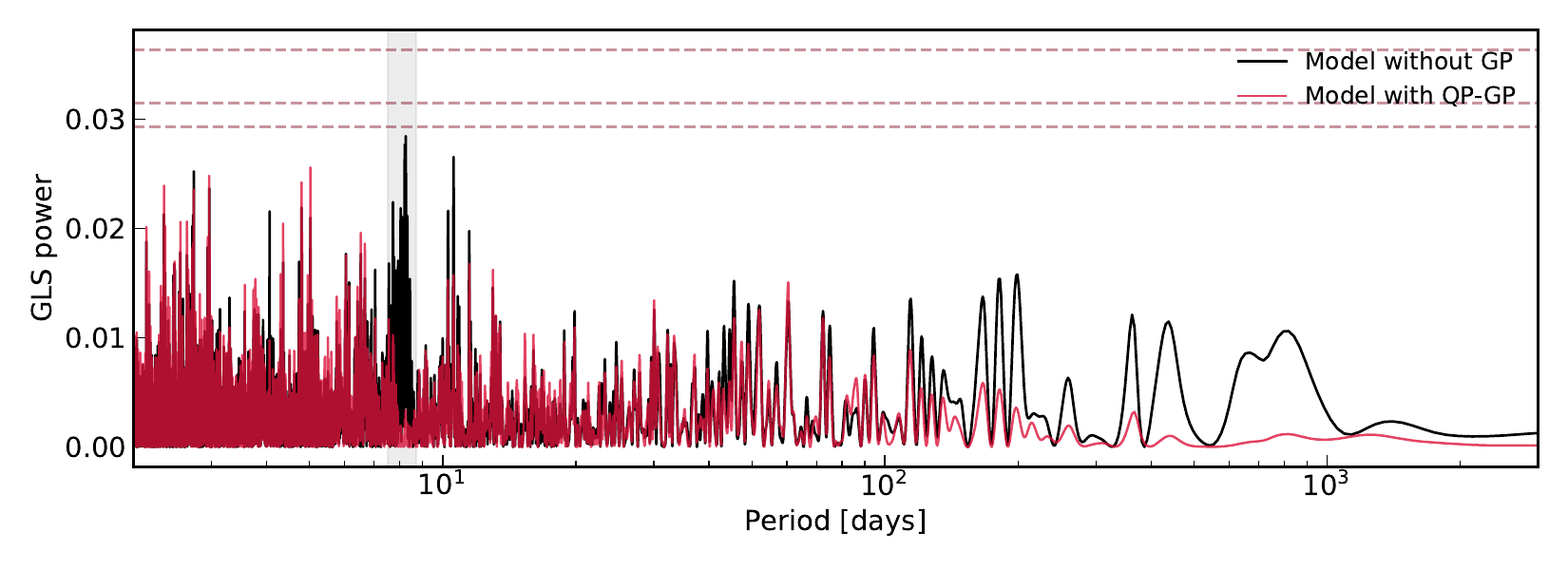}
		   \caption{Periodogram of the residuals of the RV analysis without GP (in black) and with GP (in red). The horizontal dashed lines represents the FAP levels at 0.1, 1 and 10$\%$. The shaded area correspond to the stellar rotation period.}
			  \label{fig_residuals}
\end{figure}

\section{Nested sampling analysis of planets d and e}
\label{appendice:nested_sampling}

We performed a nested sampling analysis on the HARPS, HARPS-N and ESPRESSO RV data to evaluate the orbital periods of planets d and e and to compare different models in a Bayesian framework. Nested sampling is a Monte Carlo algorithm that simultaneously explores the posterior distribution and computes the Bayesian evidence, $\mathcal{Z}$, for model comparison. We used the \texttt{jaxns} library \citep{Albert2020}, which implements a fully JAX-based nested sampling framework optimised for high-dimensional and correlated parameter spaces.

\begin{figure}[h]
		\centering
		\includegraphics[width=1\columnwidth]{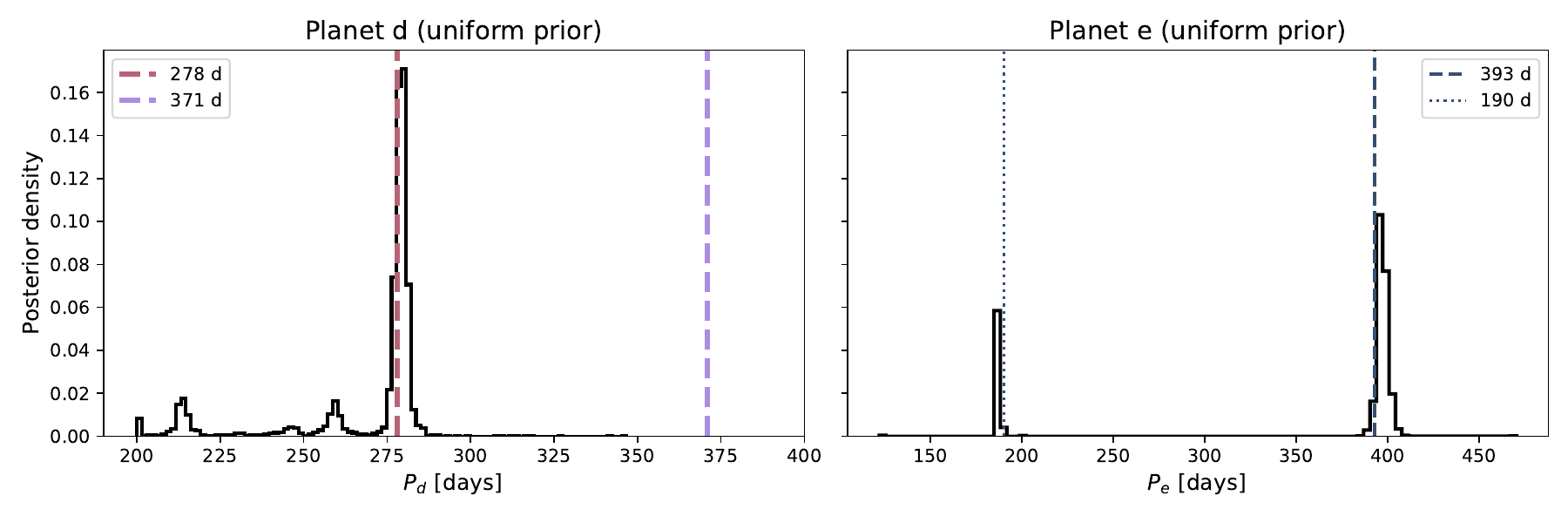}
		   \caption{Posterior distributions of the orbital periods of planets d (left) and e (right) obtained from nested sampling with uniform priors of the ESPRESSO data. The two main possible orbital periods for planet d are highlighted in pink (278 days) and purple (371 days). For planet e, the signal at $\approx 393$ days and its alias at $\approx 200$ days are highlighted in blue dashed lines. }
			  \label{fig_nested_sampling}
\end{figure}

We considered RV-only models, including the known planets b, c, g and f and the candidate planet h. We first included only planet d in order to test its orbital period. A uniform prior between 200 and 400 days was adopted to encompass the two possible periods (278 and 371 days). All other priors were identical to those used in the MCMC analysis presented in Section \ref{sec:RV}. We ran \texttt{jaxns} with 900 live points and resampled the weighted posterior to obtain 5000 effective samples. Figure~\ref{fig_nested_sampling} shows the resulting posterior distribution, which exhibits a clear preference for an orbital period of 278 days for HIP~41378~d.

We then included planet e, adopting a uniform prior between 100 and 500 days, since longer periods are unlikely given the observed transit duration. An orbital period of $P_e \approx 393$ days is preferred. An alias of this period is also present in the posterior distribution, near 190 days.

In order to constrain the orbital period of planet e, which is the most challenging to determine, we performed a nested sampling analysis combining transit data (the two K2 light curves) and RV data from the four spectrographs. We included the fit of all the other planets of the system with gaussian prior on their orbital periods. We tested three scenarios: (1) planet e located between planets d and f (uniform prior between 300 and 500 days), (2) planet e located exterior to planet f (uniform prior between 550 and 900 days), and (3) planet e located interior to planet d (uniform prior between 100 and 200 days). For each scenario, we ran \texttt{jaxns} with 900 live points.

The Bayesian evidence allows us to quantify the relative support for models with different assumptions on the orbital period of planet e. Table~\ref{tab:evidence_nested_sampling} summarises the results. The quantity $\Delta \log Z$ is computed relative to the model with the strongest Bayesian evidence; differences larger than 5 are generally interpreted as strong evidence against the disfavored model. The configuration in which planet e lies between planets d and f is strongly favoured, with a difference in Bayesian evidence of $\Delta \log Z \approx 15$ compared to the alternative configuration in which planet e is located exterior to planet f.

\begin{table}
\centering
\caption{Comparison of the tested periods for HIP 41378 e}
\label{tab:evidence_nested_sampling}
\begin{tabular}{lccc}
\hline
 Planet order & $\log Z$ & $\Delta \log Z$  & ESS \\
\hline
 $d - e - f$ & $8855.93 \pm 0.40$ & 0.0 &  5242 \\
$e - d -f$ & $8849.43 \pm 0.40$ & $-6.50$ &  4960 \\
$d - f - e$ & $8841.23 \pm 0.38$ & $-14.70$ &  4803 \\
\hline
\end{tabular}
\tablefoot{Comparison of the tested orbital periods for HIP 41378 e with photometry and RV data. The columns report, from left to right: the assumed planetary ordering, the Bayesian evidence ($\log Z$) with its uncertainty, the evidence difference $\Delta \log Z$ relative to the highest-evidence model, and the effective sample size (ESS).}
\end{table}

\section{Detection limit}
\label{appendice:detection_limit}
We computed the RV detection limit using the method introduced in \citet{Standing2022} and implemented with the diffusive nested sampler kima. The analysis was performed on the RV residuals obtained after subtracting the median best-fit signals of the seven known planets and the quasi-periodic GP.

We modelled the residuals assuming a single additional Keplerian signal using the \textit{known object} mode of kima, which samples the posterior distribution of all signals compatible with the data but not yet detected. The orbital period was assigned a log-uniform prior between 3 and $10^5$ days, while the RV semi-amplitude was drawn from a uniform prior between 0.07 and 100~\ms. The eccentricity followed a truncated normal distribution centred on zero with a dispersion of 0.083, and uniform priors were adopted for the argument of periastron and orbital phase.

To ensure a well-sampled posterior, we performed three independent runs with different random seeds, each producing more than 20,000 effective posterior samples. The resulting $(P_\mathrm{p}, K_\mathrm{p})$ posterior densities were combined and displayed as a greyscale hexbin map.

The detection limit was computed by dividing the period range into logarithmically spaced bins and, within each bin, evaluating the maximum 99th percentile of the $K_\mathrm{p}$ distribution. To mitigate small-number statistics, we required at least two posterior samples to lie above the chosen percentile in each bin. The detection limit was not computed beyond the first period bin, where the posterior became significantly affected by the upper boundary of the $K_\mathrm{p}$ prior.

\section{Tests of periodicity}
\label{appendice:asp}

 \begin{figure}[h]
		\centering
		\includegraphics[width=\columnwidth]{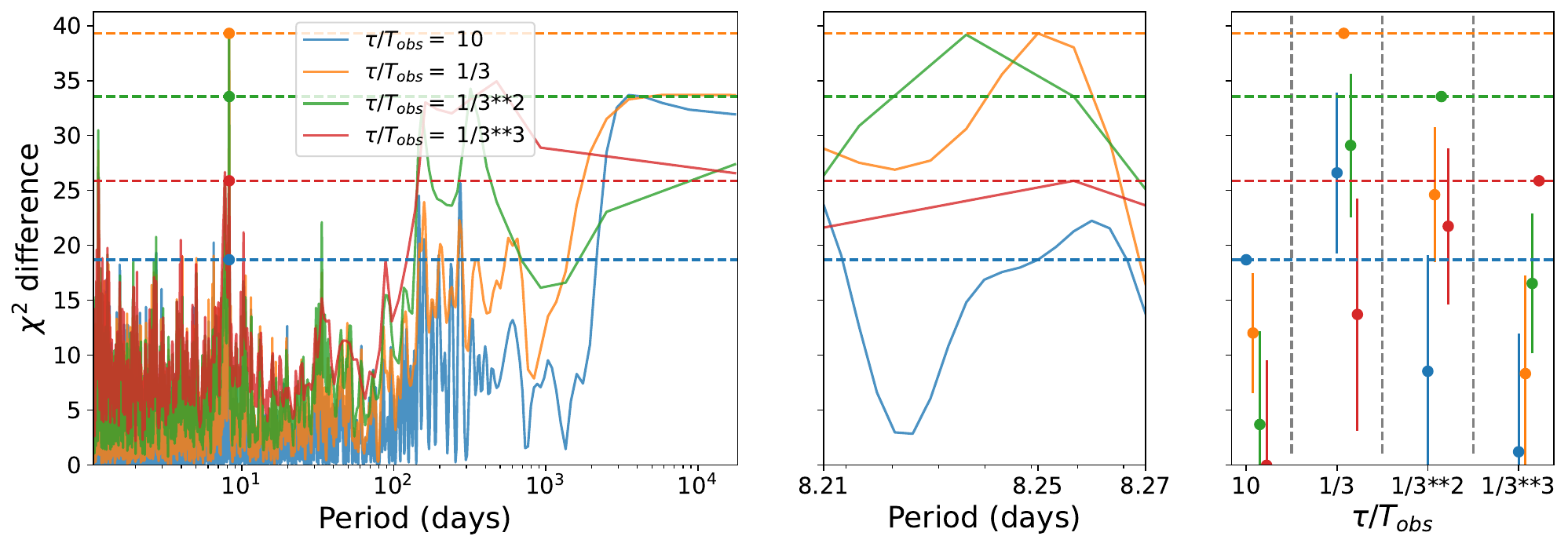}
		   \caption{Apodized sine periodogram of the RV data, including offsets, transiting planets, and order 11 polynomial in the base model. Left: apodized periodogram for different apodization timescales, that is Eq.~\eqref{eq:z} for $\tau =10T_\mathrm{obs}$ (blue), $T_\mathrm{obs}/3$ (orange), $T_\mathrm{obs}/9$ (green) and $T_\mathrm{obs}/27 $ (red), denoting by $T_\mathrm{obs}$ the observation timespan of the data. Horizontal dashed lines represent the value of the ASP at the period corresponding to the global maximum (here attained for $\tau = T_\mathrm{obs}/3$   Middle: zoom around the maximum peak. The right panel represents a statistical test: assuming the true value of the timescale given in the $x$ axis, the circles represent the expected value of the peaks corresponding to other timescales, the error bar represent the standard deviation. For instance, assuming $\tau =10T_\mathrm{obs}$ is the true timescale, we would expect an apodized sine with $\tau =T_\mathrm{obs}/3$ to have a $\chi^2$ difference of $\approx$ 12 $\pm 5$, which is 5.4 $\sigma$ away from the measured value (39, orange dashe line). The error bars are horizontally shifted to improve the readability of the figure, but they correspond to the same assumed timescale.}
			  \label{fig:asp}
\end{figure}

To assess the origin of the signals, we apply an apodized sine periodogram (ASP), which tests whether the signals have a consistent amplitude, phase, and frequency over time. It consists in computing the difference between the $\chi^2$ of a linear base model $H$, and the $\chi^2$ of a model $K(\omega, t_0,\tau)$ defined as the linear model of $H$ plus an apodized sinusoid $e^{-\frac{(t-t_0)^2}{2\tau^2}}  (A \cos \omega t + B \sin \omega t )$ \citep{Hara2022}. The ASP is defined as the $\chi^2$ difference between the two models, maximized over the center of the time window.
\begin{align}
	z(\omega, t_0,\tau) &= \max\limits_{t_0} \left(\chi^2_H - \chi^2_{K(\omega, t_0,\tau)} \right).\label{eq:z}
\end{align}

We use the same base model as in Section \ref{sec:l1}, and use four values of the signal time scale, $\tau = 10 T_\mathrm{obs}, T_\mathrm{obs}/3,T_\mathrm{obs}/9$, and $T_\mathrm{obs}/27$. The ASP is shown in Fig.~\ref{fig:asp} (left), and exhibits a strong 8.2 signal with a time scale of $\tau = T_\mathrm{obs}/3$, the centre panel is simply a zoom in.

We want to test whether the signal is statistically compatible with a constant one (i.e. with $\tau = 10 T_{obs}$). Denoting by $t_{(\tau,\omega)}$ the value of $t_0$ maximising the value of the periodogram~\eqref{eq:z} for a given $\omega$ and $\tau$, we compute the distribution of :

\begin{align}
   D_z = z(\omega, t_{(\tau,\omega)}, \tau) - z(\omega, t_{(\tau^\prime,\omega)}, \tau^\prime) 
   \label{eq:dz}
\end{align}

 with the hypothesis that the model $K(\omega, t_{(\tau,\omega)}, \tau, A^\star, B^\star )$ is correct,  where the fitted cosine and sine amplitudes $A^\star, B^\star$ are obtained by fitting model $K$ to the data. $D_z$ can easily be expressed as a generalised $\chi^2$ distribution, with mean and variance is given by an analytical expression \citep{Hara2022}. We find that a stable, periodic signal is excluded (Fig.~\ref{fig:asp}, right panel).

As a final remark, we note that the ASP is dominated by short-term signals, possibly due to systematics. Nonetheless, the two peaks at 278 and 146 days exhibited in Section~\ref{sec:l1} also appear as stable signals in Fig.~\ref{fig:asp}, left panel.

\section{Additional resonant maps}

\begin{figure}[h]
\centering
\includegraphics[width=.7\columnwidth]{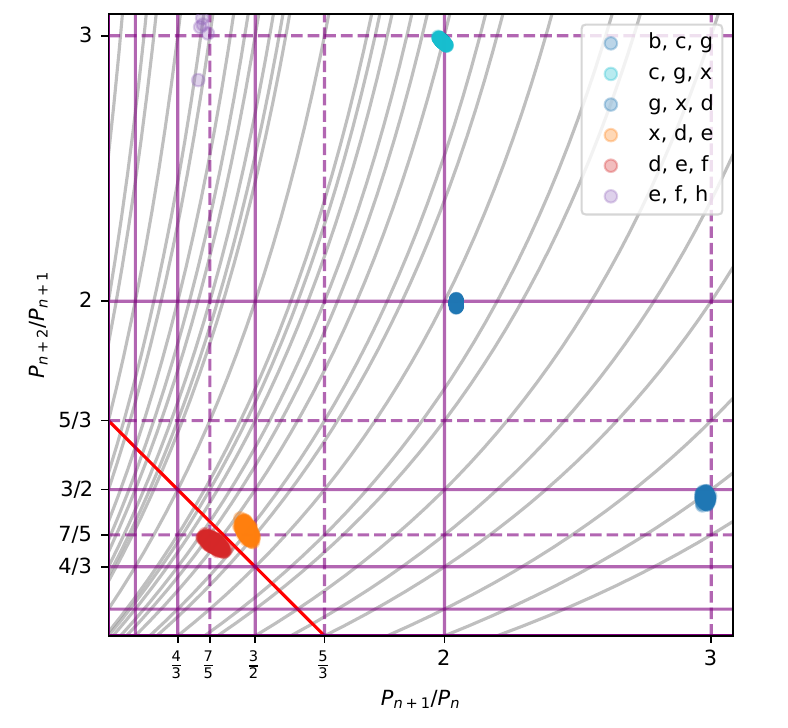}
   \caption{Position of the posterior of HIP~41378 with respect to MMR, with the addition of a planet x between g and d while minimising distance to 2-body MMRs (case i). See Figure \ref{MMR} for more details. }
      \label{MMR_p1}
\end{figure}

\begin{figure}[h]
\centering
\includegraphics[width=.7\columnwidth]{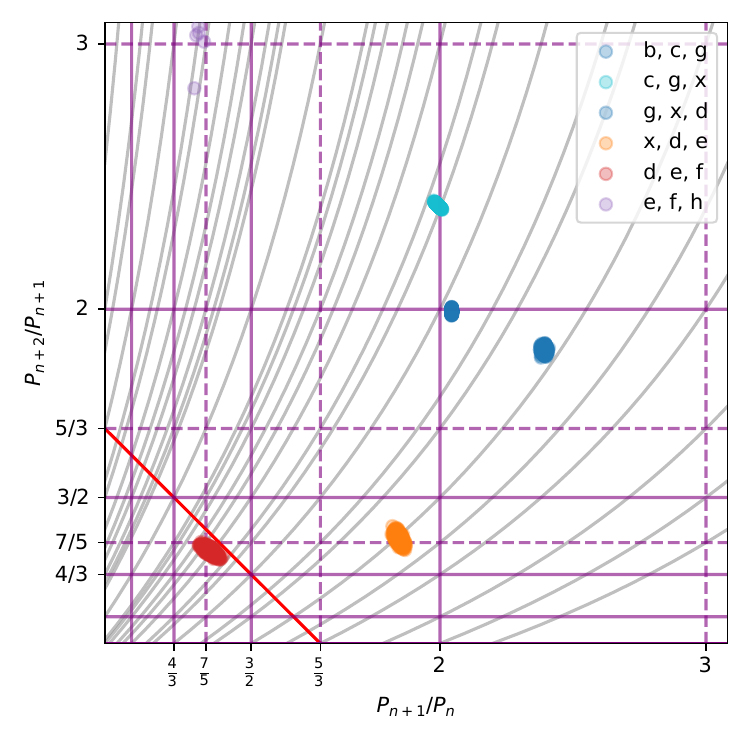}
   \caption{Position of the posterior of HIP~41378 with respect to mean motion resonance, with the addition of a planet x between g and d while minimising distance to 3-body MMRs (case ii). See Figure \ref{MMR} for more details. }
      \label{MMR_p1_3b}
\end{figure}

\onecolumn

\section{Activity Indicators periodograms}
 \begin{figure}[h]
		\centering
		\includegraphics[width=1\textwidth]{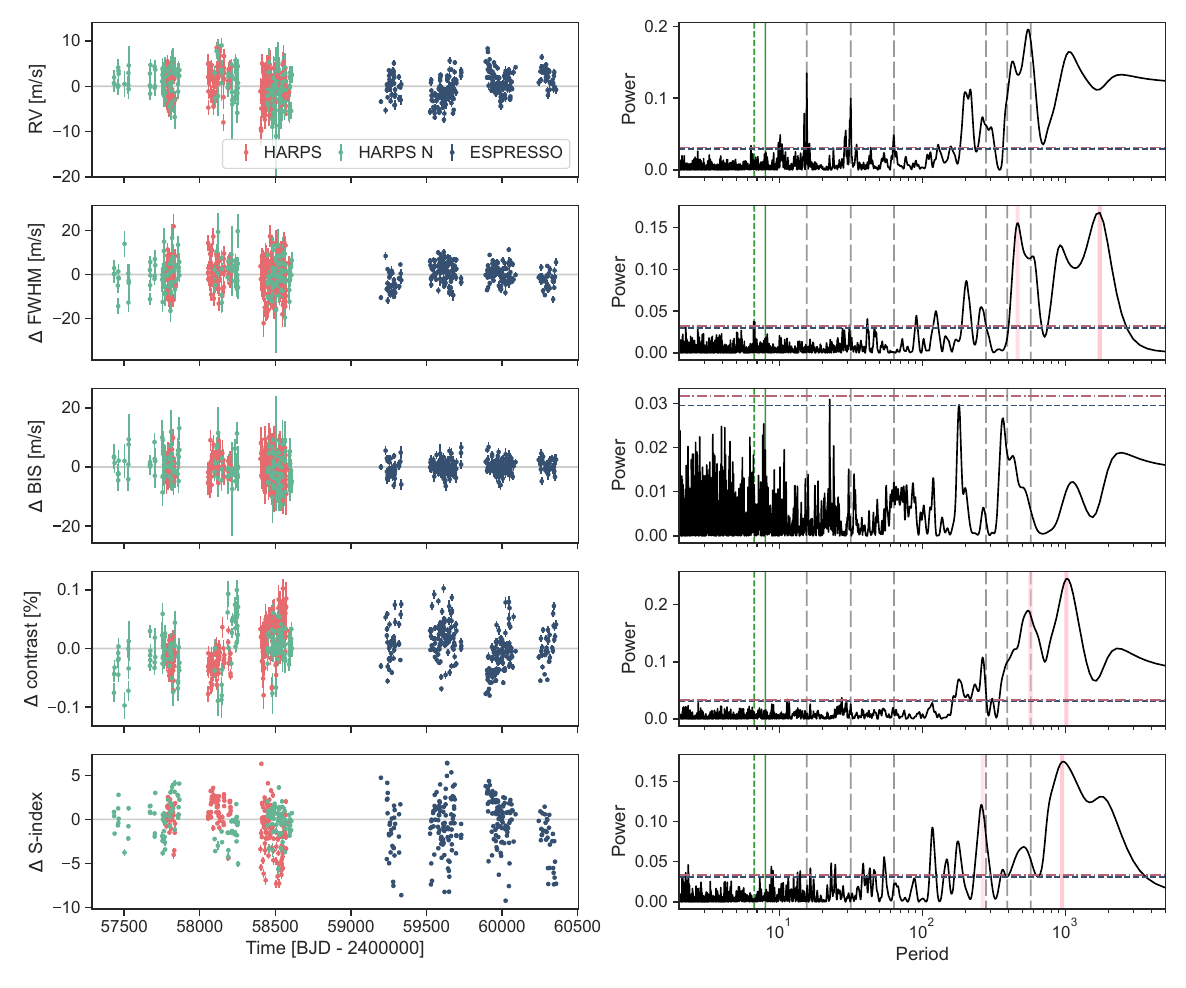}
		   \caption{Time series and GLS periodograms of RVs and stellar activity indicators for HIP~41378 from HARPS, HARPS-N, and ESPRESSO. From top to bottom: RVs, FWHM, BIS, CCF contrast, and S-index. Periods of planets b, c, g, d, e and f are shown as grey dashed lines. Stellar rotation periods from photometry and $\log(R'_\mathrm{HK})$ are indicated as green dotted and solid lines, respectively. Maximum power periods are shown as dark pink lines, with aliases in light pink. False-alarm probabilities at 1\% and 10\% are indicated with horizontal lines.} 
			  \label{fig_periodograms_activity}
\end{figure}

\newpage
\section{RV analysis priors and posteriors distributions}

 \begin{figure}[h]
		\centering
		\includegraphics[width=0.85\columnwidth]{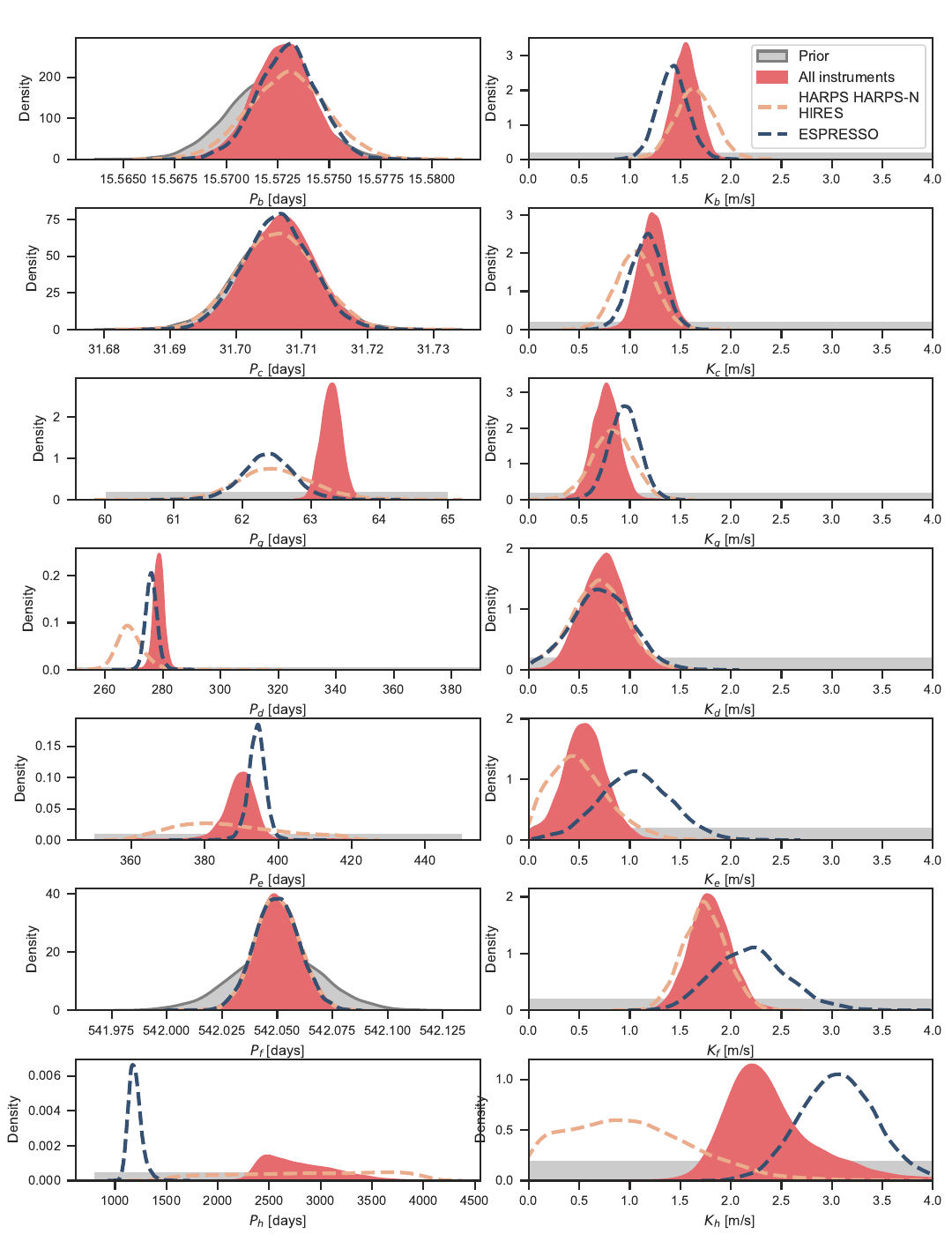}
		   \caption{Prior and posterior distributions of the orbital periods (left panels) and RV semi-amplitudes (right panels) for the seven planets. The priors are shown as grey distributions. The dashed orange lines indicate the posterior distributions obtained from the RV analysis using the HARPS, HARPS-N, and HIRES instruments. The dashed blue lines correspond to the posterior distributions derived from the ESPRESSO data alone. The pink distributions represent the posterior results from the combined analysis of all four instruments.}
			  \label{fig_posterior_priors_RV}
\end{figure}

\newpage

\section{Global analysis results}
\begin{longtable}{lcc}
\caption{List of posteriors with median and $68\%$ highest density intervals} 

\\
\hline
\textbf{Parameter} & \textbf{Prior} & \textbf{Posterior} \\
\hline
\endfirsthead

\multicolumn{3}{c}{{\bfseries \tablename\ \thetable{} -- continued from previous page}} \\
\hline
\textbf{Parameter} & \textbf{Prior} & \textbf{Posterior} \\
\hline
\endhead

\hline
\multicolumn{3}{r}{{Continued on next page}} \\
\hline
\endfoot

\hline
\multicolumn{3}{l}{\footnotesize
\textit{Note.} $\mathcal{U}$ denotes a uniform distribution, $\mathcal{N}$ a normal (Gaussian) distribution, and $\mathcal{LN}$ a log-normal distribution.}\\
\multicolumn{3}{l}{
For the log-normal prior, the distribution is defined such that $\ln x \sim \mathcal{N}(\mu,\sigma^2)$.}\\ 
\multicolumn{3}{l}{The priors for the periods of planets d and e have been chosen after model comparison.}\\ 
\multicolumn{3}{l}{The masses of planets d and e also have the 95$\%$ credible upper limit} 
\endlastfoot

\textbf{Planet b}  &    &                \\
Orbital period P$_b$ [days]   &  $\mathcal{N}(15.572, 0.002)$  & $15.57209 \pm 0.00002$       \\
RV semi-amplitude K$_b$ [\ms]    &   $\mathcal{U}(0, 10)$   &  $1.56_{-0.10}^{+0.12}$     \\
Mid-transit time T$_{0,b}$ [BJD]   &   $\mathcal{N}(2457152.2844, 0.0021)$      & $2457152.282 \pm 0.0019$        \\
Eccentricity $e_b$      &      $\mathcal{N}_{[0,1]}(0.0, 0.083)$         & $0.050^{+0.027}_{-0.050}$  \\
Argument of periastron $\omega_b$ [rad]      &      $\mathcal{U}(-\pi, \pi)$        &  $-1.26^{+0.89}_{-0.88}$  \\
Cos inclination $\cos i_b$      &      $\mathcal{U}(-\pi, \pi)$        &  $0.02013 \pm 0.00084$ \\
Transit depth $\delta F_b$      &  $\mathcal{LN}$(ln($3\times10^{-4}$),0.5)            &  $3.44\times10^{-4} \; ^{+3.4\times10^{-6}}_{-4.1\times10^{-6}}$ \\
Planet mass M$_b$ [\Mearth]   &    derived        & $6.90^{+0.48}_{-0.50} $  \\

Planet-to-star radius ratio $R_{p,b}/R_*$   &    derived & $0.01854 \; ^{+9\times10^{-5}}_{-1.1\times10^{-4}}$  \\
Planet radius $R_b$ [\Rearth]   &    derived &  $2.64^{+0.022}_{-0.024}$\\
Orbital inclination $i_b$  [deg]          &     derived            & $88.847 \pm +0.048$ \\
Impact parameter b$_b$           &     derived            & $ 0.448 \pm 0.029$ \\
Transit duration T$_{14,b}$ [days]   &  derived   & $5.26  \pm 0.21$ \\
Density $\rho_b$ [\gcm] & derived & $2.07 \pm 0.16$\\
Semi-major axis a$_b$ [AU] & derived & $0.1300 \pm 0.0009$\\
\hline

\textbf{Planet c}  &    &                \\
Orbital period P$_c$ [days]   &  $\mathcal{N}(31.706, 0.006)$  & $31.70591 \pm 0.00005$       \\
RV semi-amplitude K$_c$ [\ms]    &   $\mathcal{U}(0, 10)$   &  $1.06^{+0.12}_{-0.13}$     \\
Mid-transit time T$_{0,c}$ [BJD]   &   $\mathcal{N}(2457163.1659, 0.0027)$      & $2457163.166^{+0.0016}_{-0.0013}$        \\
Eccentricity $e_c$      &      $\mathcal{N}_{[0,1]}(0.0, 0.083)$        & $0.044^{+0.022}_{-0.044}$  \\
Argument of periastron $\omega_c$ [rad]      &      $\mathcal{U}(-\pi, \pi)$       &  $1.10^{+2.04}_{-2.32}$ \\
Cos inclination $\cos i_c$      &      $\mathcal{U}(-\pi, \pi)$        &   $0.02661 \pm 0.00022$\\
Transit depth $\delta F_c$      &   $\mathcal{LN}$(ln($3\times10^{-4}$),0.5)            &  $3.70\times10^{-4} \; ^{+1.1\times10^{-5}}_{-1.2\times10^{-5}}$ \\
Planet mass M$_b$ [\Mearth]   &    derived        & $5.97^{+0.56}_{-0.86}$   \\
Planet-to-star radius ratio $R_{p,c}/R_*$   &    derived & $0.01923 \; ^{+2.9\times10^{-4}}_{-3.1\times10^{-4}}$  \\
Planet radius $R_c$ [\Rearth]   &    derived &  $2.74^{+0.043}_{-0.049}$\\

Orbital inclination $i_c$  [deg]          &     derived            & $88.475 \pm 0.013$ \\
Impact parameter b$_c$           &     derived            & $0.910^{+0.036}_{-0.045}$ \\
Transit duration T$_{14,c}$ [days]   &  derived   & $3.22^{+0.49}_{-0.30}$ \\
Density $\rho_c$ [\gcm] & derived & $1.60 \pm 0.21$\\
Semi-major axis a$_c$ [AU] & derived & $0.2088 \pm 0.0014$\\

\hline
\textbf{Planet g}  &    &                \\
Orbital period P$_g$ [days]   &  $\mathcal{U}(60, 65)$  &    $63.19 \pm 0.12$  \\
RV semi-amplitude K$_g$ [\ms]    &   $\mathcal{U}(0, 10)$   & $0.81^{+0.10}_{-0.12}$   \\
Orbital phase $\Phi_g$ & $\mathcal{U}(-0.5, 0.5)$ & $-0.33 \pm 0.03$\\
Eccentricity $e_g$      &      $\mathcal{N}_{[0,1]}(0.0, 0.083)$         & $0.055^{+0.024}_{-0.054}$  \\
Argument of periastron $\omega_g$ [rad]      &      $\mathcal{U}(-\pi, \pi)$        &  $0.21^{+2.92}_{-1.49}$ \\
Mid-transit time t$_{0,g}$ [BJD]   &   derived      &   $2458777.86^{+1.86}_{-1.82}$      \\
Planet minimum mass M$_g$ [\Mearth]   &    derived        & $5.75^{+0.76}_{-0.83}$   \\
Semi-major axis a$_g$ [AU] & derived & $0.3307 \pm 0.0023$\\
\hline

\textbf{Planet d}  &    &                \\
Orbital period P$_d$ [days]   &  $\mathcal{N}(278.3616, 0.5)$  &  $278.3618 \pm 0.0003$  \\
RV semi-amplitude K$_d$ [\ms]    &   $\mathcal{U}(0, 10)$   &  $0.56^{+0.23}_{-0.27}$     \\
Mid-transit time T$_{0,d}$ [BJD]   &   $\mathcal{N}(2457166.2604, 0.0007)$      &   $2457166.261 ^{+0.00061}_{-0.00057}$     \\
Eccentricity $e_d$      &      $\mathcal{N}_{[0,1]}(0.0, 0.083)$         & $0.063^{+0.027}_{-0.063}$  \\
Argument of periastron $\omega_d$ [rad]      &      $\mathcal{U}(-\pi, \pi)$        &  $-0.79^{+2.00}_{-2.35}$ \\
Cos inclination $\cos i_d$      &      $\mathcal{U}(-\pi, \pi)$        &  $0.00368 \pm -0.00009$ \\
Transit depth $\delta F_d$      &   $\mathcal{LN}$(ln($7\times10^{-4}$),0.5)            &  $6.62\times10^{-4} \; ^{+4.4\times10^{-6}}_{-5.4\times10^{-6}}$ \\
Planet mass M$_d$ [\Mearth]   &    derived        & $6.53^{+2.65}_{-3.15}$ \textbf{(<11.65)}   \\
Planet-to-star radius ratio $R_{p,d}/R_*$   &    derived & $0.02573 \; ^{+9\times10^{-5}}_{-1.1\times10^{-4}}$ \\
Planet radius $R_d$ [\Rearth]   &    derived &  $3.65 \pm 0.029$\\
Orbital inclination $i_d$  [deg]          &     derived            & $89.789 \pm 0.005$ \\
Impact parameter b$_d$           &     derived            & $0.54 \pm 0.03$ \\
Transit duration T$_{14,d}$ [days]   &  derived   & $12.71^{+0.46}_{-0.39}$ \\
Density $\rho_d$ [\gcm] & derived & $0.74 \pm 0.33$\\
Semi-major axis a$_d$ [AU] & derived & $0.8885^{+0.0060}_{-0.0058}$ \\
\hline

\textbf{Planet e}  &    &                \\
Orbital period P$_e$ [days]   &  $\mathcal{U}(387, 450)$  &  $393_{-5}^{+3}$  \\
RV semi-amplitude K$_e$ [\ms]    &   $\mathcal{U}(0, 10)$   &  $0.58^{+0.24}_{-0.36}$     \\
Mid-transit time T$_{0,e}$ [BJD]   &   $\mathcal{N}(2457142.01954, 0.0007)$      &    $2457142.018^{+0.00051}_{-0.00048}$    \\
Eccentricity $e_e$      &      $\mathcal{N}_{[0,1]}(0.0, 0.083)$         & $0.065_{+0.031}^{-0.065}$  \\
Argument of periastron $\omega_e$ [rad]      &      $\mathcal{U}(-\pi, \pi)$        &  $0.82^{+2.32}_{-1.53}$ \\
Cos inclination $\cos i_e$      &      $\mathcal{U}(-\pi, \pi)$        &   $0.00348 \pm 0.00006$\\
Transit depth $\delta F_e$      &   $\mathcal{LN}$(ln($1\times10^{-3}$),0.5)            &  $1.33\times10^{-3} \; ^{+1.03\times10^{-5}}_{-1.07\times10^{-5}}$ \\
Planet mass M$_e$ [\Mearth]   &    derived        & $7.62^{+3.20}_{-4.63}$ \textbf{(<15.66)}   \\
Planet-to-star radius ratio $R_{p,e}/R_*$   &    derived & $0.03648 \pm 1.4\times10^{-4}$  \\
Planet radius $R_e$ [\Rearth]   &    derived &  $5.19 \pm 0.04$\\
Orbital inclination $i_e$  [deg]          &     derived            & $89.801 \pm 0.003$ \\
Impact parameter b$_e$           &     derived            & $0.632^{+0.034}_{-0.039}$ \\
Transit duration T$_{14,e}$ [days]   &  derived   & $13.20^{+0.33}_{-0.26}$ \\
Density $\rho_e$ [\gcm] & derived & $0.30^{+0.13}_{-0.18}$\\
Semi-major axis a$_e$ [AU] & derived & $1.1195^{+0.0094}_{-0.011}$\\
\hline

\textbf{Planet f}  &    &                \\
Orbital period P$_f$ [days]   &  $\mathcal{N}(542.07975, 0.0014)$  &  $542.0797 \pm 0.0001$  \\
RV semi-amplitude K$_f$ [\ms]    &   $\mathcal{U}(0, 10)$   &  $1.78^{+0.35}_{-0.38}$     \\
Mid-transit time T$_{0,f}$ [BJD]   &   $\mathcal{N}(2457186.9145, 0.0003)$      &   $2457186.914^{+0.00018}_{-0.00022}$     \\
Eccentricity $e_f$      &      $\mathcal{N}_{[0,1]}(0.0, 0.083)$         & $0.052^{+0.025}_{-0.052}$  \\
Argument of periastron $\omega_f$ [rad]      &      $\mathcal{U}(-\pi, \pi)$        &  $-0.16^{+1.87}_{-1.46}$ \\
Cos inclination $\cos i_f$      &      $\mathcal{U}(-\pi, \pi)$        &  $0.00090 \pm 0.00013$ \\
Transit depth $\delta F_f$      &  $\mathcal{LN}$(ln($4\times10^{-3}$),0.5)             &  $4.43\times10^{-3} \; ^{+1.42\times10^{-5}}_{-1.49\times10^{-5}}$ \\
Planet mass M$_f$ [\Mearth]   &    derived        & $25 \pm 5$   \\
Planet-to-star radius ratio $R_{p,f}/R_*$   &    derived & $0.06656 \pm 1.1\times10^{-4}$  \\
Planet radius $R_f$ [\Rearth]   &    derived &  $9.47 \pm 0.07$\\
Orbital inclination $i_f$  [deg]          &     derived            & $ 89.948 \pm 0.008$ \\
Impact parameter b$_f$           &     derived            & $0.207^{+0.034}_{-0.031}$ \\
Transit duration T$_{14,f}$ [days]   &  derived   & $18.9^{+0.85}_{-0.96}$ \\
Density $\rho_f$ [\gcm] & derived & $0.166 \; ^{+0.033}_{-0.036}$\\
Semi-major axis a$_f$ [AU] & derived & $1.3856^{+0.0094}_{-0.0090}$\\
\hline

\textbf{Planet h (candidate)}  &    &                \\
Orbital period P$_h$ [days]   &  $\mathcal{U}(800, 4000)$  &  $2602_{-433}^{+468}$    \\
RV semi-amplitude K$_h$ [\ms]    &   $\mathcal{U}(0, 10)$   &  $1.79^{+0.61}_{-0.55}$  \\
Orbital phase $\Phi_h$ & $\mathcal{U}(-0.5, 0.5)$ & $-0.40 \pm 0.09$\\
Eccentricity $e_h$      &      $\mathcal{N}_{[0,1]}(0.0, 0.083)$         & $0.066^{+0.033}_{-0.066}$  \\
Argument of periastron $\omega_h$ [rad]      &      $\mathcal{U}(-\pi, \pi)$        &  $-0.73^{+1.01}_{-2.39}$ \\
Mid-transit time T$_{0,h}$ [BJD]   &   derived      &   $2458291^{+239}_{-225}$      \\
Planet minimum mass M$_h$ [\Mearth]   &    derived        & $43^{+16}_{-13}$   \\
Semi-major axis a$_f$ [AU] & derived & $3.94^{+0.41}_{-0.51}$\\
\hline
\textbf{Gaussian Process}  &    &                \\
GP period P$_{rot}$ [days] &  & $8.160^{+0.111}_{-0.071}$\\
$l$ [TBD] & $\mathcal{U}(1.0, 500)$ & $100^{+60}_{-70}$\\
$\gamma$ [TBD] & $\mathcal{U}(0.1, 10)$ & $0.94^{+0.52}_{-0.83}$\\
Amplitude $A$ [\ms] & $\mathcal{U}(0.01, 5.0)$ & $1.28^{+0.22}_{-0.28}$\\
\hline
\textbf{Instruments}  &    &                \\
offset$_{HARPS}$  &   $\mathcal{N}(0, 2)$          & $0.03^{+0.55}_{-0.47}$  \\
offset$_{HARPS-N}$ &  $\mathcal{N}(0, 2)$          & $-0.30^{+0.51}_{-0.48}$   \\ 
offset$_{HIRES}$ &  $\mathcal{N}(0, 2)$          & $-0.22^{+0.57}_{-0.58}$   \\ 
offset$_{ESPRESSO}$ &   $\mathcal{N}(0, 2)$      & $-0.06^{+0.61}_{-0.56}$   \\
$\sigma_{HARPS}$  &   $\mathcal{H}(2)$          & $1.48^{+0.19}_{-0.21}$  \\
$\sigma_{HARPS-N}$ &  $\mathcal{H}(2)$          & $1.64^{+0.27}_{-0.23}$   \\ 
$\sigma_{HIRES}$ &   $\mathcal{H}(2)$      & $3.66^{+0.28}_{-0.30}$   \\
$\sigma_{ESPRESSO}$ &   $\mathcal{H}(2)$      & $1.424^{+0.083}_{-0.084}$   \\

\label{posteriors}
\end{longtable}

\section{K2 transit light curves}

 \begin{figure}[h]
		\centering
		\includegraphics[width=1.0\columnwidth]{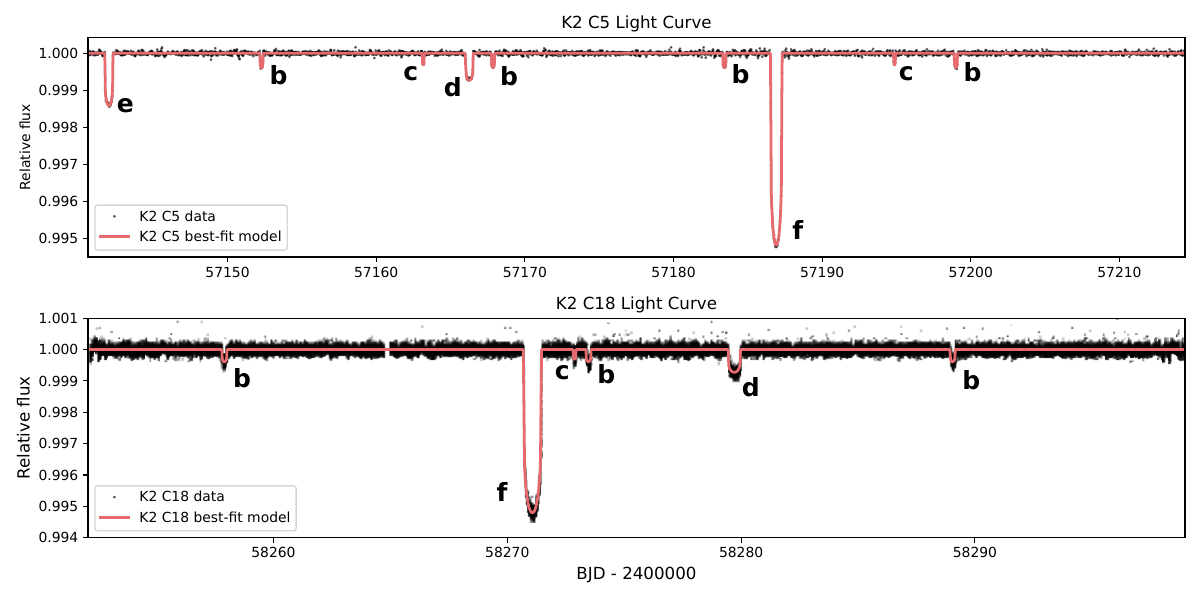}
		   \caption{Light curves of HIP 41378. \textit{Upper panel}: K2 C5 campaign (long cadence). \textit{Bottom panel}: K2 C18 campaign (short cadence). The pink curve represents the median of the best-fit model from photometry and RV analysis. The names of the planets are highlighted with their letters.}
			  \label{fig_fit_transit}
\end{figure}

\end{appendix}

\end{document}